\documentclass[twocolumn]{aastex63}

\usepackage{epstopdf}
\usepackage{multimedia}
\usepackage{epsfig,graphics,subfigure,psfrag,amsmath,amssymb,amscd}
\usepackage{url}

\begin{document}

\title{A possible tidal disruption event candidate in the black hole binary system of OJ 287}

\author{Shifeng Huang}
\affiliation{Shandong Key Laboratory of Optical Astronomy and Solar-Terrestrial Environment,\\
 School of Space Science and Physics, \\
 Institute of Space Sciences, Shandong University,  \\
 Weihai, Shandong, 264209, China.
 \href{Corresponding author.}{husm@sdu.edu.cn, yinhx@sdu.edu.cn}}

\author{Shaoming Hu}
\affiliation{Shandong Key Laboratory of Optical Astronomy and Solar-Terrestrial Environment,\\
 School of Space Science and Physics, \\
 Institute of Space Sciences, Shandong University,  \\
 Weihai, Shandong, 264209, China.
 \href{Corresponding author.}{husm@sdu.edu.cn, yinhx@sdu.edu.cn}}

\author{Hongxing Yin}
\affiliation{Shandong Key Laboratory of Optical Astronomy and Solar-Terrestrial Environment,\\
 School of Space Science and Physics, \\
 Institute of Space Sciences, Shandong University,  \\
 Weihai, Shandong, 264209, China.
 \href{Corresponding author.}{husm@sdu.edu.cn, yinhx@sdu.edu.cn}}

\author{Xu Chen}
\affiliation{Shandong Key Laboratory of Optical Astronomy and Solar-Terrestrial Environment,\\
 School of Space Science and Physics, \\
 Institute of Space Sciences, Shandong University,  \\
 Weihai, Shandong, 264209, China.
 \href{Corresponding author.}{husm@sdu.edu.cn, yinhx@sdu.edu.cn}}

\author{Sofya Alexeeva}
\altaffiliation{LAMOST fellow}
\affiliation{CAS Key Laboratory of Optical Astronomy, National Astronomical Observatories, Beijing, 100101, China.
 }

\author{Dongyang Gao}
\affiliation{Shandong Key Laboratory of Optical Astronomy and Solar-Terrestrial Environment,\\
 School of Space Science and Physics, \\
 Institute of Space Sciences, Shandong University,  \\
 Weihai, Shandong, 264209, China.
 \href{Corresponding author.}{husm@sdu.edu.cn, yinhx@sdu.edu.cn}}
\affiliation{Key Laboratory of Modern Astronomy and Astrophysics,  Nanjing University,\\
Ministry of Education, Nanjing 210023, China}

\author{Yunguo Jiang}
\affiliation{Shandong Key Laboratory of Optical Astronomy and Solar-Terrestrial Environment,\\
 School of Space Science and Physics, \\
 Institute of Space Sciences, Shandong University,  \\
 Weihai, Shandong, 264209, China.
 \href{Corresponding author.}{husm@sdu.edu.cn, yinhx@sdu.edu.cn}}

\begin{abstract}
 The BL Lacertae OJ 287 is a supermassive black holes binary (SMBHB) system with complex physics of its irregular flares.
 During 2016 October -- 2017 April period, a surprising outburst in the X-ray, UV and optical bands was detected, while no variability was seen in the $\gamma$-ray light curve. During the outburst, the X-ray light curves were dominated by the soft X-rays, whose peak in the luminosity was $\sim$ $10^{46}\text{erg}~\text{s}^{-1}$ --more than 10 times higher than the mean level before the outburst--and a ``softer-when-brighter" phenomenon was exhibited. These above phenomena have been reported by some previous works.
The hardness ratio showed negligible evolution with flare time and soft X-ray luminosity. Critically, the luminosity of the soft X-rays decayed following a power law of $t^{-5/3}$ which occurs in most tidal disruption events (TDEs), and similar trend can be seen in UV and optical bands during the soft X-ray declining period. Helium and oxygen narrow emission lines
are strengthened prominently in the optical spectra of postoutburst epochs, that could be attributed to the surrounding gas appeared due to TDE. We discuss three possible origins of the event, including the jet's precession, the aftereffect of the black hole-disk impaction and the TDE. Our results show that the TDE is the more likely scenario to explain
the outburst.

\end{abstract}


\section{Introduction}

The nearby OJ 287 ($z$ = 0.306) has attracted people for many years due to its roughly 12 yr quasiperiodic outburst in the optical light curve, but a great breakthrough occurred at the end of the 1980s when \cite{sillanpaa1988} proposed the object to be a supermassive black holes binary (SMBHB).
To explain the unique light curve, \cite{lehto1996} proposed the central engine model.
 According to their model, a smaller black hole crosses the accretion disc of a larger black hole during the binary orbit of the black holes about each other. The model can explain the double-peak structure in about every 12 yr of the light curve and some optical outbursts were predicted successfully \citep{valtonen2008,valtonen2016,laine2020}. However, the origin of the variability of the light curve is so complex that one can not predict all outbursts with a single scenario. The disk precession \citep{katz1997}, beaming effect \citep{villata1998}, the precession of the jet \citep{abraham2000,britzen2018} and periodically tidal disruption by SMBHB system \citep{tanaka2013} were proposed to explain the quasiperiodic outburst of OJ 287. Some optical outbursts can also be explained by the tidally induced by the secondary black hole \citep{valtonen2009}. These models mainly concern the optical outburst, especially the roughly 12 yr quasiperiod. However, for other phenomena in light curves, such as the variation in the X-ray light curve, the above models are not so plausible.

When a star occasionally passes by the tidal radius of a black hole, it will be captured and torn into debris if the self-gravity can not overcome the tidal force. This phenomenon is called a tidal disruption event (TDE). During the TDE, roughly half of the debris is bound to the black hole, resulting in a rapidly rising luminosity and furthering its power-law decay $t^{-5/3}$ \citep{rees1988,phinney1989}.

During the flare, a soft X-ray outburst can be observed  \citep{bade1996,komossa1999a,greiner2000,maksym2014,komossa2015,saxton2015,holoien2019} with the contribution of radiation from the disk and the wind \citep{strubbe2009,lodato2009,lodato2011,metzger2016}. In TDEs, periods including the Lense-Thiring period and Kepler period can be found with X-ray data \citep{reis2012,pasham2019,miniutti2019}. Several nonjetted TDE candidates were found
in inactive galaxies \citep{bade1996,komossa1999a,alexander2016}, SMBHBs \citep{saxton2012,liu2014,shu2020}, and active galactic nuclei (AGNs;  \citealt{komossa1999a,grupe2015,campana2015,blanchard2017,ricci2020}).
 Besides, the jet signatures were detected in three TDE candidates, including Swift J164449.3+573451 \citep{bloom2011,burrows2011,levan2011,zauderer2011,kara2016}, Swift J2058.4+0516 \citep{cenko2012}, and Swift J1112.2-8238 \citep{brown2015} through the radio, hard X-ray and $\gamma$-ray observations.
The TDE rate in a single black hole galaxy is expected to be $10^{-5}-10^{-4}~\text{yr}^{-1} \text{galaxy}^{-1}$ \citep{rees1990,velzen2014}, while in AGNs or SMBHBs, the rate can be much higher \citep{karas2007,chen2009,komossa2015}. In typical TDEs, broad hydrogen and helium emission lines
can be seen in the spectra \citep{gezari2012,macleod2012,arcavi2014,bogdanovic2014,komossa2015} and
these spectroscopic signatures depend on the property of the progenitor of the disrupted star.

The X-ray luminosity of OJ 287 was at a low level for a long time until 2016 October, when the multiwavelength outburst was observed \citep{komossa2017,kapanadze2018,kushwaha2018,komossa2020}. A strong correlation between the X-ray, optical, and UV bands in the outburst epoch was  detected, while no variability was seen in the $\gamma$-ray and radio bands  \citep{kushwaha2018}. During the  outburst, the soft X-ray luminosity reached an extreme higher value compared to other epochs. On the contrary, the hard X-ray did not rise significantly, and the hardness ratio (HR) was detected
at the lowest level in the \emph{Swift} monitor record. The X-ray energy spectrum analysis yielded a ``softer-when-brighter" phenomenon \citep{komossa2017,komossa2020,komossa2021}. According to \cite{kushwaha2018}, the strong X-ray flare emitted from the new component was related to the jet activity. Besides, \cite{kapanadze2018} believed that the radiation was related to the shock in the jet and its precession contributed to the X-ray outburst.

In this work, we present our investigations of the sudden soft X-ray flare of OJ 287 during 2016 -- 2017 period employing the light curves and spectra.
Three scenarios including the jet precession, aftereffect of the black hole-disk impaction and the TDE are considered when we investigate the origin of the outburst. In section 2, the data reduction processes are introduced. Then, the multiwavelength light curves and polarization degree, the X-ray light curves and hardness, the spectra of X-ray and optical are present in section 3. In addition, to investigate the origin of the outburst, three models are discussed in section 4.
Finally, the conclusion is derived in section 5.
The cosmological parameters, $H_0=69.3~\text{km}~\text{s}^{-1}\text{Mpc}^{-1}$, $\Omega_M=0.287$, and $\Omega_{\Lambda}=0.713$,
were adopted from the Wilkinson Microwave Anisotropy Probe's ninth-year results (WMAP9) \citep{hinshaw2013}.

\section{Observation and Data Reduction}
\subsection{\emph{Fermi $\gamma$-ray data}}
The main instrument on the Fermi Gamma-ray Space Telescope mission, the Large Area Telescope (\emph{Fermi}-LAT), is a high-energy $\gamma$-ray telescope monitoring the energy range from 20 MeV to 300 GeV across the sky \citep{atwood2009}. We downloaded the $\gamma$-ray data of OJ 287 from the \emph{Fermi}-LAT data website\footnote{\url{https://fermi.gsfc.nasa.gov/cgi-bin/ssc/LAT/LATDataQuery.cgi}} for the period from 2016 January 1 to 2019 September 1.

The data were analyzed with standard unbinned-likelihood tutorials by using the Fermi Science Tools \texttt{Fermitools} version 1.0.10. The event data were extracted by \texttt{gtselect} while selecting the source region centered on the coordinates of the OJ 287 with the radius of the search region set at $10^{\circ}$. The good time intervals were selected with the task \texttt{gtmktime}. A counts map of the degree region of interest (ROI) was created by \texttt{gtbin} and the exposure map was generated by \texttt{gtltcube} and \texttt{gtexpmap}. With the background models, including the current Galactic diffuse emission model \rm{gll\_iem\_v07.fits} and the corresponding model for the extragalactic isotropic diffuse emission iso\_P8R3\_SOURCE\_V2\_v1.txt, the XML files were created by \rm{make4FGLxml.py}, while and the diffuse source responses were created by \texttt{gtdiffrsp}. Finally, the task \texttt{gtlike} was run to obtain the flux.

\subsection{\emph{Swift XRT and UVOT observations}}
\subsubsection{\emph{X-ray data}}
The \emph{Swift} X-Ray Telescope (XRT) is one of the instruments on the Neil Gehrels \emph{Swift} observatory with a sensitive broad-band detector for X-ray from 0.3 to 10 keV \citep{burrows2005}. The data from 2007 March 6 to 2019 June 13 were retrieved from the High Energy Astrophysics Science Archive Research Center (HEASARC) website\footnote{\url{https://heasarc.gsfc.nasa.gov/db-perl/W3Browse/w3browse.pl}} and we
followed standard threads\footnote{\url{https://www.swift.ac.uk/analysis/xrt/index.php}} to analyse the data from level I.

All \emph{Swift} data were reduced with \texttt{HEASoft 6.26.1}.
The task \texttt{xrtpipeline} was run, with a source region file selected by a central circle with radius of $47^{\prime\prime}$ and a background region file selected for an annulus with an inner radius of $165^{\prime\prime}$ and outer radius of $235^{\prime\prime}$. The light curves and spectra were generated by \texttt{xselect} with level II data from the photon-counting (PC) mode and windowed-timing (WT) mode, respectively.

The source spectra were grouped by \texttt{grppha} with a minimum of 20 photons $\text{bin}^{-1}$ for WT mode spectra and at least 5 photons $\text{bin}^{-1}$ for PC mode spectra. In the reduction of X-ray spectra, the response matrix file swxwt0to2s6\_20131212v015.rmf for WT mode and swxpc0to12s6\_20130101v014.rmf for PC mode were used and the standard ancillary response files were created by \texttt{xrtmkarf}. To improve the signal-to-noise ratio (S/N), five spectra which were collected before the outburst were joined as the preoutburst spectrum, and six postoutburst spectra were combined as the post-outburst spectrum. Specifically, the spectra with observed IDs 00030901184, 00030901185, 00030901187, 00030901188 and 00030901189 were extracted to combine as the preoutburst spectrum, while the outburst spectrum was extracted with observed ID 00033756114, and the postoutburst spectrum was combined from the spectra extracted from observed IDs 00034934053, 00034934054, 00034934055, 00034934056, 00034934057 and 00034934058.

\subsubsection{\emph{UV-optical data}}
The UV/Optical Telescope (UVOT) has an effective wavelength range  from 170-600 nm employs 7 filters (\textsl{v}, \textsl{b}, \textsl{u}, \textsl{uvw1}, \textsl{uvm2}, \textsl{uvw2} and white), a field of view $17^{\prime} \times 17^{\prime}$, and is one of instruments on the \emph{Swift} observatory \citep{roming2005}. Following the recommended threads\footnote{\url{https://www.swift.ac.uk/analysis/uvot/index.php}}, \texttt{uvotsource} in \texttt{HEASoft 6.26.1} was run with a source circular region radius of $10^{\prime\prime}$ with a background region with the circular radius of $50^{\prime\prime}$ for the \textsl{v}, \textsl{b}, and \textsl{u} bands, but for \textsl{uvw1}, \textsl{uvw2} and \textsl{uvm2} we used a source region file with radius $15^{\prime\prime}$ and a background file with a $75^{\prime\prime}$ circular radius. We adopted a Galactic extinction of 0.11, 0.15, 0.18,
0.24, 0.34, and 0.32 mag in the \textsl{v}, \textsl{b}, \textsl{u}, \textsl{uvw1}, \textsl{uvm2} and \textsl{uvw2} bands, respectively, following to \cite{kapanadze2018}.

\subsection{\emph{Ground based optical and radio light curves}}
The 1 m Cassegrain telescope is located at Weihai Observatory of Shandong University \citep{hu2014}. The data during the periods from 2014 March 15 to 2014 August 4, from 2016 October 31 to 2016 December 29,  and from 2018 December 31 to 2019 June 30 were recorded by the PI PIXIS 2048B CCD, the rest of the data were taken by the Andor DZ936 CCD. Johnson-Cousins \textsl{UBVRI} filters were used for the observations. All obtained data were reduced automatically with an IDL script \citep{chen2014} and then they were calibrated  using standard stars \citep{fiorucci1996}.

The data from Steward Observatory, University of Arizona, were obtained with 2.3 m Bok Telescope and 1.54 m Kuiper Telescope with the SPOL CCD Imaging/Spectropolarimeter \citep{smith2009}. The polarimetry data at the wavelength range 5000--7000 {\AA} obtained from 2008 October to 2018 June were downloaded from the public website\footnote{\url{http://james.as.arizona.edu/~psmith/Fermi/}}.

The history light curve for \textsl{R} band range from 2005 to 2012 was extracted from \cite{nilsson2018}.

For the \textsl{V}, \textsl{R} and \textsl{I} bands, the magnitudes were corrected with Galactic extinctions of 0.11 mag for the \textsl{V} band, 0.06 mag for the \textsl{R} band and 0.04 mag for the \textsl{I} band \citep{schlafly2011,kapanadze2018}.
The magnitudes were converted to fluxes following to \cite{bessell1998}.

The 40 m telescope at the Owens Valley Radio Observatory (OVRO), monitors more than 1800 blazars \citep{richards2011}. The 15 GHz data covering from 2016 January to 2019 May for OJ 287 were obtained from their public data archive website\footnote{\url{http://www.astro.caltech.edu/ovroblazars/index.php?page=home}}.

\subsection{\emph{Optical spectra data}}
Optical spectra from the Sloan Digital Sky Survey (SDSS) DR16\footnote{\url{https://www.sdss.org/dr16/}} and the Large Sky Area Multi-Object Fiber Spectroscopic Telescope (LAMOST) DR6\footnote{\url{http://dr6.lamost.org/}} were analyzed.  We obtained two pre-outburst optical spectra from LAMOST DR6, but unfortunately, due to the low S/N, the only spectrum that can be used is the one obtained in 2016 January. Another optical spectrum for the preoutburst phase was observed in 2005 October and we extracted it from SDSS. Unfortunately, neither SDSS nor LAMOST observed OJ 287 during the outburst interval, hence we only extracted  spectra which were observed after the outburst. To enhance the S/N, two spectra obtained on 2017 December 18 and 2017 December 21 were joined together. Three spectra obtained on 2018 January 8, 11 and 12 were also joined for analysis.

\section{Results}
\subsection{\emph{Multiwavelength light curves during outburst}}
\begin{figure}
\begin{center}
\includegraphics[width=0.5\textwidth]{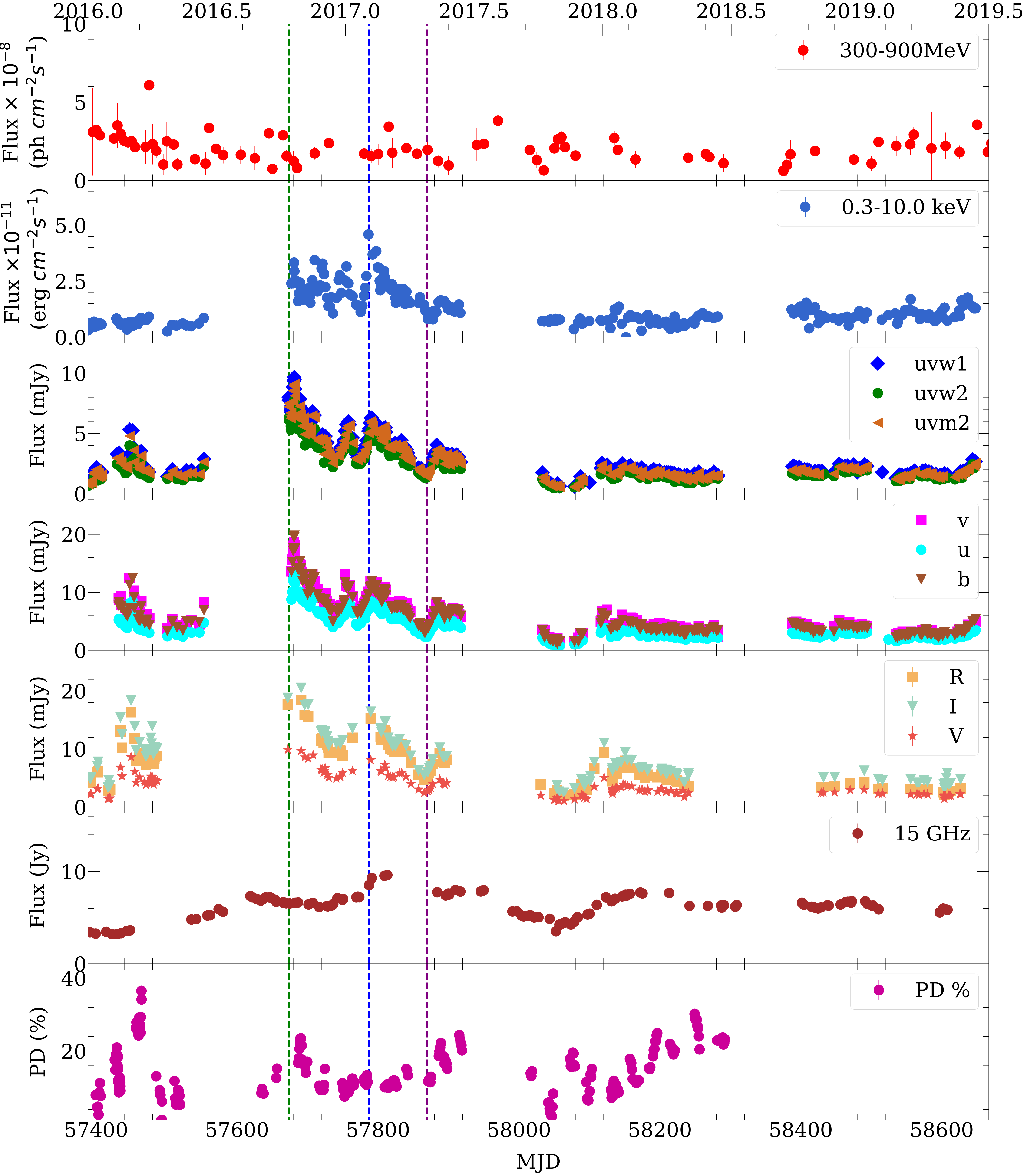}
\end{center}
\caption{Multiwavelength light curves for OJ 287 from 2016 January to  2019 June.
The data in the first panel were extracted from \emph{Fermi}-LAT, the data in the second panel are from \emph{Swift/XRT}, the data in the third and the fourth panels are from \emph{Swift/UVOT}, the data in the fifth panel are from Weihai Observatory, the radio data in the sixth panel are from OVRO, and the data of optical polarization degree (PD) ranging from $5000\text{\AA}$ to $7000\text{\AA}$ on the bottom panel are from Steward Observatory. The dashed lines in each panel corresponding to different epochs during the X-ray outburst: the green one marks the {\it Swift/XRT} WT mode start time, the blue one notes the time of the X-ray peak and the purple one is at the time of the end of the WT mode observations.}
\label{fig:mwlc}
\end{figure}

The multiwavelength light curves of OJ 287 during 2016 January--2019 June, ranging from the $\gamma$-ray to the radio, are shown in Figure \ref{fig:mwlc}. Three vertical dashed lines mark the important epochs of the outburst: the green one marks the time when \emph{Swift/XRT} began to observe in WT mode, the blue one dashed line corresponds to the peak level of the X-ray light curve and the purple one illustrates the end of the observations with WT mode during the outburst. It should be noted that in this interval, the light curves ranging from the X-ray to optical bands manifest rapidly rise, although the $\gamma$-ray one does not. It is difficult to determine the the start time of the outburst due to the small separation between OJ 287 and the Sun in the summer of 2016, so few data were collected then.
On MJD 57681 (where MJD = JD $-$ 2400000.5) the UV and optical bands reached the peak levels for their light curves but the X-ray light curve did not peak then. For the UV and optical bands, the light curves show a declining trend during MJD 57681--57869, with a higher polarization degree ($23.51\pm 0.05\%$) at the beginning of that interval. During MJD 57673 -- 57775  period, the X-rays display a variable flux but still keep to a level which is much higher than before;  in the interval MJD 57775--57786, the flux rapidly rises roughly 4 times in 11 days reaching the peak in MJD 57786. It should be noted that, when the X-ray flux reaches the maximum value, the polarization degree has declined to a relatively lower level ($11.53 \pm 0.05 \%$). The radio light curve rises to a high level during MJD 57673--57936, but very smoothly, and shows a similar tendency with the UV, optical and X-rays in the interval MJD 57786--57869, while in the range MJD 57673--57786 it does not.  During this time, the $\gamma$-ray flux shows a slight variation but no clear correlation with the other bands.

\subsection{\emph{X-ray light curve and HR}}

\begin{figure}
\begin{center}
\includegraphics[width=0.5\textwidth]{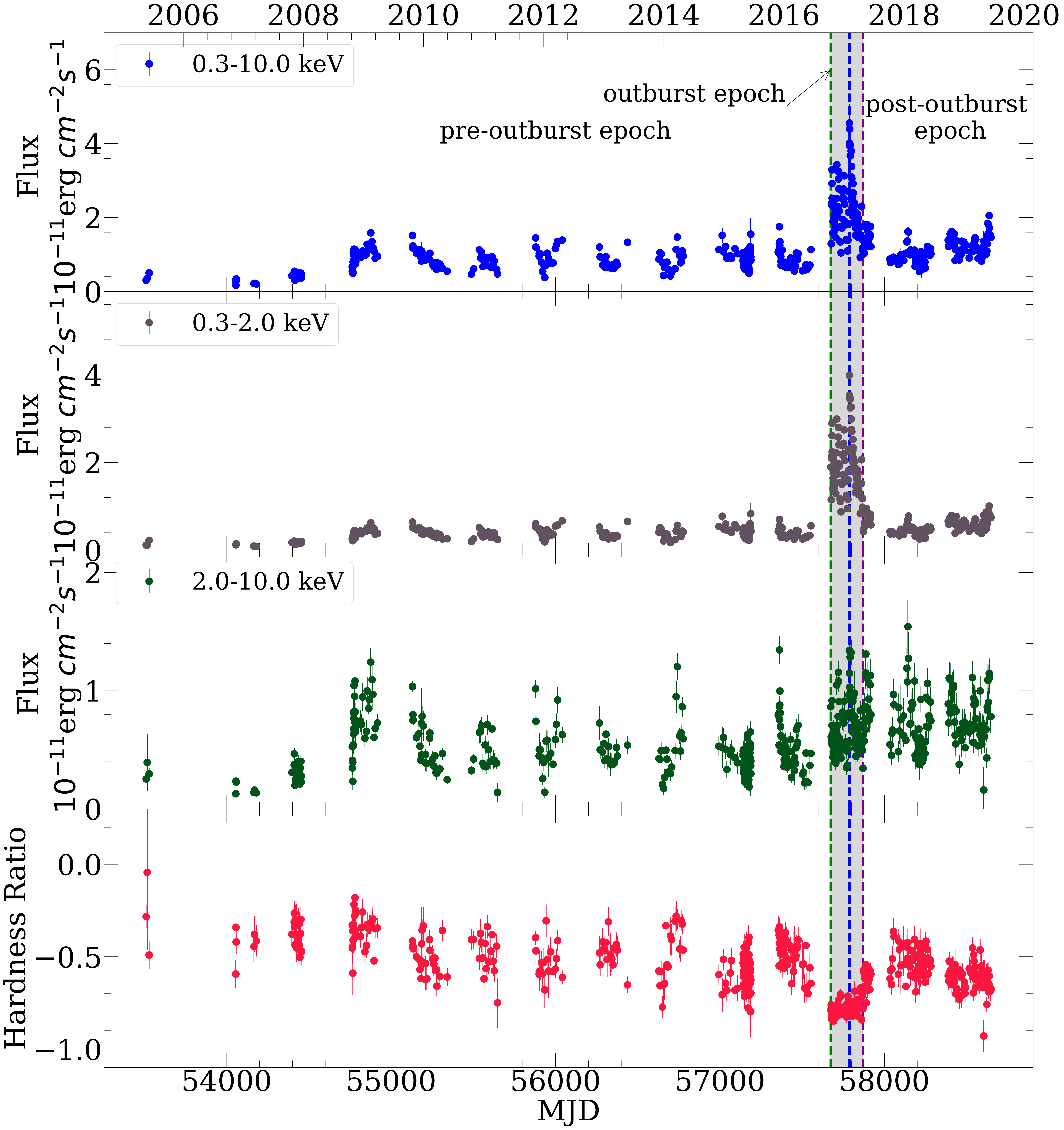}
\end{center}
\caption{The X-ray light curves and HR for OJ 287 during \emph{Swift} monitoring. The four panels (top to bottom) are the light curves of 0.3--10.0 keV (total), 0.3--2.0 keV (soft), 2.0--10.0 keV (hard), and the HR, respectively. The grey shaded region in each panel marks the outburst epoch. The three dashed lines in each panel mark the outburst start, peak, and end times, respectively.}
\label{fig:xlc}
\end{figure}

In this subsection, we show the X-ray light curves since OJ 287 since 2005, when \emph{Swift} began monitoring OJ 287. The HR is defined by the count rate ratio
\begin{equation}
HR=\frac{CR(2.0\mbox{--}10.0 \text{keV})-CR(0.3\mbox{--}2.0 \text{keV})}{CR(0.3\mbox{--}2.0 \text{keV})+CR(2.0\mbox{--}10.0 \text{keV})},
\end{equation}
where CR(2.0--10.0 keV) and CR(0.3--2.0 keV) are the count rates of the hard X-ray and soft X-ray, respectively. Here we define the MJD 54763--57553 as the preoutburst epoch, and the postoutburst epoch is defined as MJD 57869--58648.

In Figure \ref{fig:xlc}, the light curves reveal a prominent outburst of soft X-ray during 2016 October -- 2017 April (MJD 57673 -- 57870).
The hard X-ray flux has a high level in \emph{Swift} monitoring history, but not higher than in postoutburst epoch. During X-ray outburst, the HR declines to the lowest level have ever been detected in \emph{Swift} monitoring history. The mean energy flux of the soft X-rays during the outburst, is $1.932 \times 10^{-11}\text{erg}~\text{cm}^{-2} \text{s}^{-1}$, while in the preoutburst epoch, the mean energy flux is $0.376 \times 10^{-11}\text{erg}~\text{cm}^{-2} \text{s}^{-1}$. For the hard X-rays, the energy flux is $0.673 \times 10^{-11}\text{erg}~\text{cm}^{-2} \text{s}^{-1}$ during the outburst and $0.484 \times 10^{-11}\text{erg}~\text{cm}^{-2} \text{s}^{-1}$ in the pre-outburst epoch. The mean HR is $-0.785\pm 0.002$ for the outburst interval and $-0.497 \pm 0.005$ for pre-outburst epoch.

\begin{figure}
  \centering
  \includegraphics[width=0.45\textwidth]{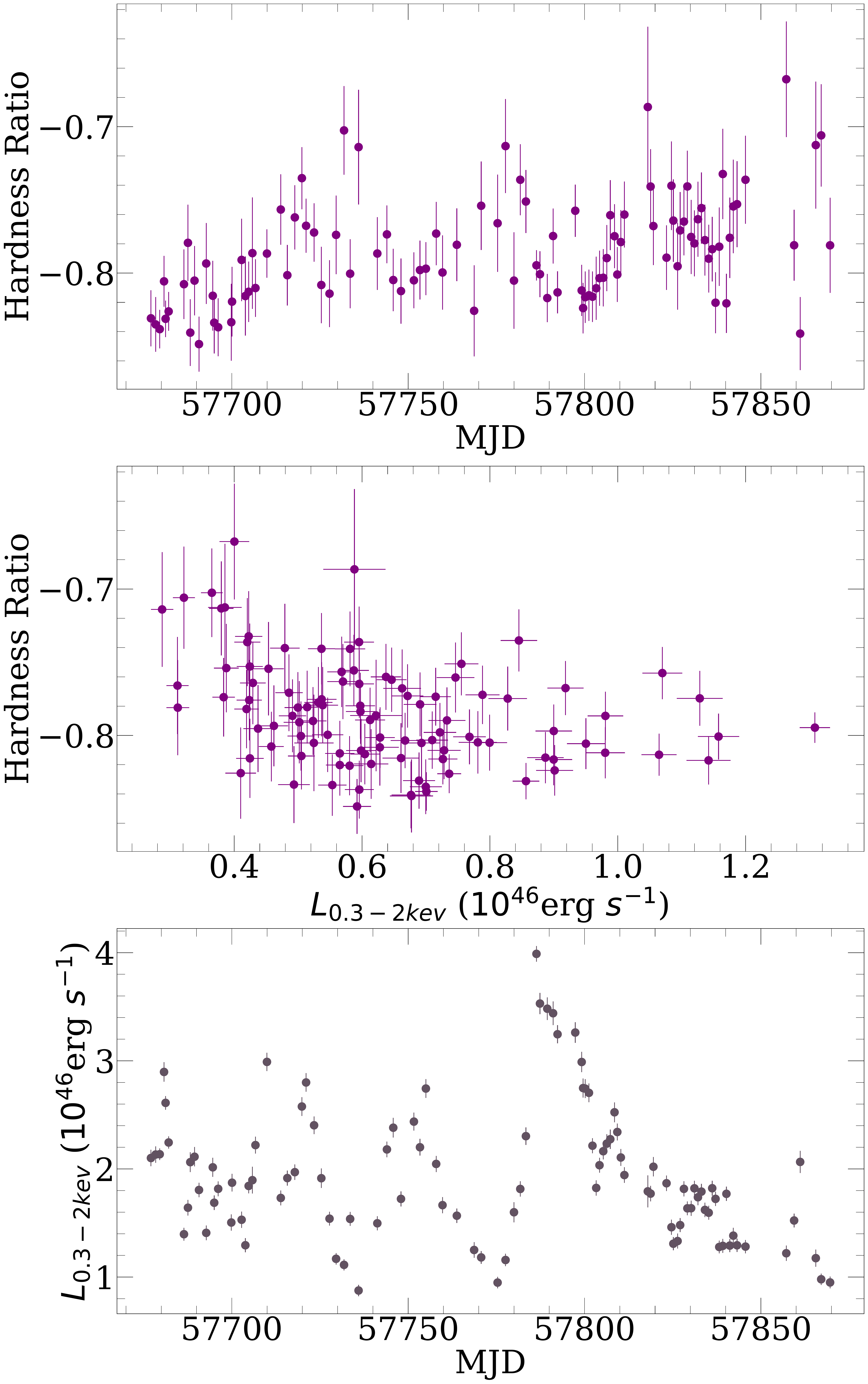}
  \caption{Evolution of the HR and soft X-ray during the outburst epoch.
The correlations between HR and time (top panel) and the correlation between HR and the soft X-ray (0.3--2.0 keV) luminosity (middle panel) are presented. In the bottom panel, the evolution of the soft X-ray luminosity is plotted. }\label{fig:hr_evo}
\end{figure}

The variation of HR in the outburst epoch is shown in Figure \ref{fig:hr_evo}. During the outburst, the HR was located at a very low level (HR ranging from $-0.849\pm 0.019$ to $-0.668\pm0.040$) and slightly evolved with time. In addition, with soft X-ray luminosity increase, the HR did not show significant variation.

\subsection{\emph{The energy spectra of X-rays}}

Fixing the neutral hydrogen column density ($\text{N}_\text{H}$), the grouped spectra are fitted with the package \texttt{xspec 12.10.1}. Three spectra from the preoutburst, outburst and postoutburst were fitted by a red shifted power law (zPL), red shifted log-parabolic (zLP) and double red shift power law (DzPL). To avoid the indetermination in the low energy of the WT mode data, the energy range of the spectra is limited to 0.5--10.0 keV in the fitting. It should be noted that in the spectral fitting, the C-statistic is carried out.

We attempted to add a disk component to the zPL or zLP models (i.e.~$tbabs*(diskbb+zpowerlw)$ or $tbabs*(diskbb+zlogpar)$) but unfortunately, the fitting results did not improve prominently. The results in the Table \ref{tab:spec_fit} reveal that during the outburst, the spectrum tends to be softer than the spectra of the preoutburst and postoutburst epochs (softer-when-brighter). We assume that the observed outburst X-ray is constructed by an existing component and the new one, therefore, the model DzPL ($tbabs*(zpowerlw+zpowerlw)$) was applied in X-ray spectra fitting. Here, in the fitting, the coefficients in the first component are fixed. It should be noted that the coefficients in the fixed component are extracted from the fitting results of a single zPL model for the preoutburst epoch spectrum. The new component may exist as the remnant in the postoutburst epoch, with this consideration, the DzPL model also can be applied to the postoutburst spectra.

\begin{figure}
  \centering
  \subfigure [pre-outburst]{
  \includegraphics[angle=90,scale=.3,trim={0 0 40 0},clip]{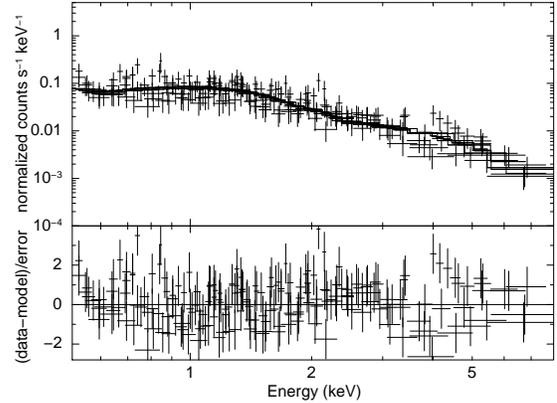}}
  \subfigure [Outburst]{
  \includegraphics[angle=90,scale=.3, trim={0 0 40 0},clip]{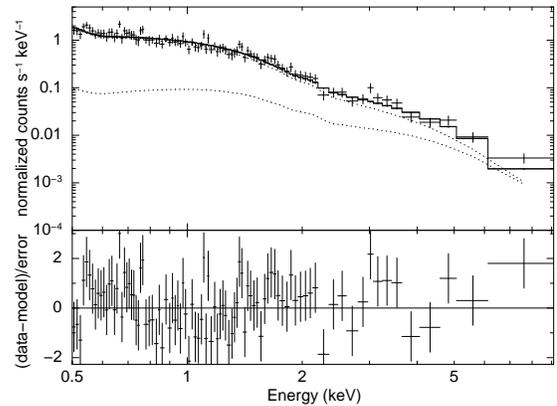}}
  \subfigure [post-outburst]{
  \includegraphics[angle=90,scale=.3, trim={0 0 40 0},clip]{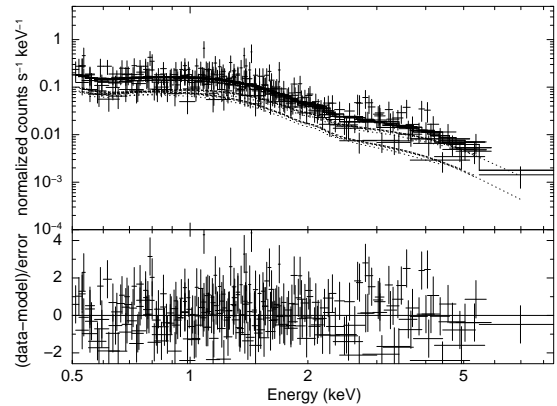}}
  \caption{Best fits of the X-ray energy spectra. Panels (a), (b), and (c) are the results of the spectra from preoutburst, outburst and postoutburst periods, respectively. Panel (a) is fitted by the model zPL. Panels (b) and (c) show the results fitted by the model DzPL. The dotted lines mark the contribution of the individual components to the fits. To improve the quality of the fits, 5 spectra collected in the preoutburst epoch are joined together, and six spectra are joined for the postoutburst epoch.}
  \label{fig:xspec_fit}
\end{figure}

\begin{deluxetable*}{lcccc}
\tablenum{1}
\tablecaption{The X-ray spectra fitted results\label{tab:spec_fit}}
\tablewidth{0pt}
\tabletypesize{\scriptsize}
\tablehead{
\colhead{Models} &  \colhead{Parameter} & \colhead{Pre-outburst} & \colhead{Outburst} & \colhead{Post-outburst}\\
}
\decimalcolnumbers
\startdata
  & $\Gamma$ & $1.75^{+0.09}_{-0.08}$ & $2.65^{+0.06}_{-0.06}$ & $1.98^{+0.08}_{-0.08}$ \\
$\text{zPL}(\texttt{tbabs}\ast \texttt{zpowerlw})$ & $K_P(10^{-3}$  $\text{ph}/\text{keV}/\text{cm}^2/\text{s})$ &$1.62^{+0.13}_{-0.12}$ & $22.76^{+0.94}_{-0.92}$ & $3.31^{+0.21}_{-0.20}$ \\
& $\chi^2_r$/(dof) & $1.28/(194)$ & $1.24/(92)$ & $1.25/(257)$ \\
\hline
 & a & $1.91^{+0.23}_{-0.23}$ & $2.90^{+0.12}_{-0.12}$ & $1.86^{+0.19}_{-0.20}$\\
$\text{zLP}(\texttt{tbabs}\ast \texttt{zlogpar})$ & b & $-0.2^{+0.29}_{-0.28}$ & $-0.45^{+0.19}_{-0.19}$ & $0.19^{+0.28}_{-0.28}$ \\
& $K_L(10^{-3}$ $\text{ph}/\text{keV}/\text{cm}^2/\text{s})$ & $1.67^{+0.14}_{-0.15}$ & $22.90^{+0.94}_{-0.91}$ & $3.25^{+0.23}_{-0.22}$ \\
& $\chi_r^2$/(dof) & $1.29/(193)$ & $1.12/(91)$ & $1.25/(256)$ \\
\hline
& $\Gamma_2$ & - & $2.82_{-0.08}^{+0.08}$ & $2.27_{-0.18}^{+0.20}$\\
$\text{DzPL}(\texttt{tbabs}\ast (\texttt{zpowerlw+zpowerlw}))$ & $K_{D2}(10^{-3}$ $\text{ph}/\text{keV}/\text{cm}^2/\text{s})$ & - & $21.34_{-0.93}^{+0.96}$ & $1.68_{-0.21}^{+0.22}$ \\
& $\chi^2_r$/(dof) & - & $1.16/(92)$ & $1.27/(257)$ \\
\enddata
\tablecomments{Considering the interstellar medium absorption model $\texttt{tbabs}$ in the fitting, the hydrogen column density was adopted as $N_H=2.49\times 10^{20}\text{cm}^{-2}$ \citep{kapanadze2018} in these fits. Here $\chi^2_r$ and $dof$ denote the reduced $\chi^2$ and degrees of freedom, respectively. $\Gamma$ and $K_P$ denote the photon index and normalization factor in zPL ($\texttt{zpowerlw}$) fits and $a$, $b$ and $K_L$ are, respectively, the photon index, curvature term and normalization factor in zLP ($\texttt{zlogpar}$) fits. For the DzPL ($\texttt{zpowerlw+zpowerlw}$) fits $\Gamma_2$ and $K_{D2}$ are the index and normalization factor for the second component, respectively; for the first component of DzPL fits, we fix the index $\Gamma_1=1.75$ and the normalization factor $K_{D1}=1.62 \times 10^{-3} \text{ph}~\text{keV}^{-1} \text{cm}^{-2}~\text{s}^{-1}$ so we only show the parameter fits to the second component fitted.}
\end{deluxetable*}

The fitting results of the three models are shown in Table \ref{tab:spec_fit}. The outburst spectrum is not fitted so well by the zPL with the reduced $\chi^{2}$ = 1.24, while for zLP and DzPL, it is 1.12 and 1.16, respectively. The reduced $\chi^{2}$ of DzPL is higher than that of zLP, but inconspicuous. Work by \cite{kushwaha2018} revealed that it was dominated by a new component during the outburst. Therefore, DzPL describes the new component very naturally, and we argue that the model is the better one. The X-ray spectral fitting results can be seen in Figure \ref{fig:xspec_fit}.

\begin{figure}
  \centering
  \includegraphics[width=0.5\textwidth]{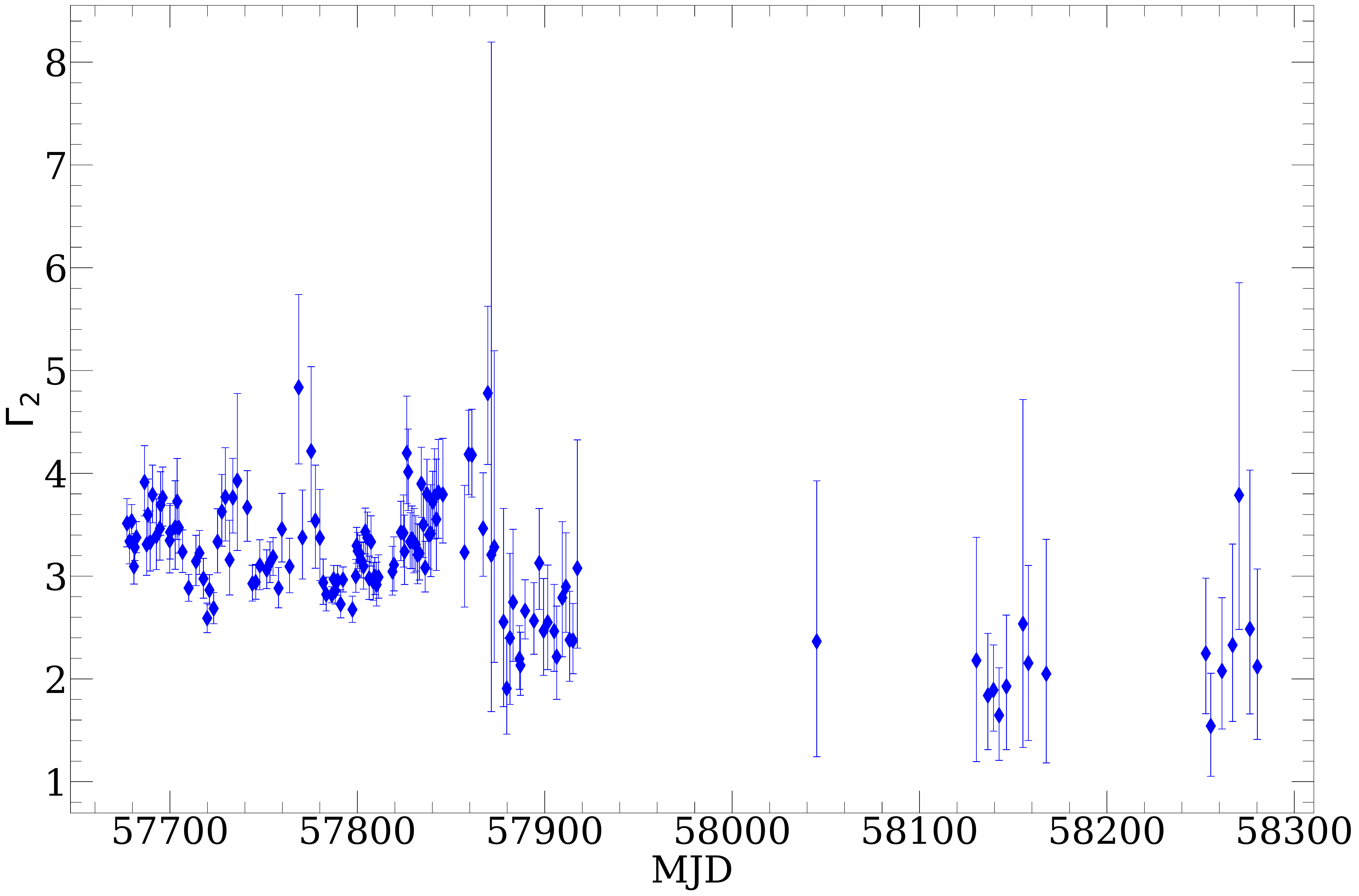}
  \caption{Temporal evolution of $\Gamma_2$. The plotted points are obtained through selecting in the results of the X-ray spectral fitting subject to the constraints that the exposure time is $>$ 500s and the degree of freedom is $>$ 5. }\label{fig:index}
\end{figure}

It is assumed that the evolution of the new component manifests as the temporal variation of the spectral index. Furthermore, the variation of the index of the second component can be seen in Figure \ref{fig:index}, which reveals that there is a very soft component during the outburst and postoutburst epochs.

\subsection{\emph{The evolution of optical spectra}}
\begin{figure}
  \centering
  \includegraphics[width=0.47\textwidth]{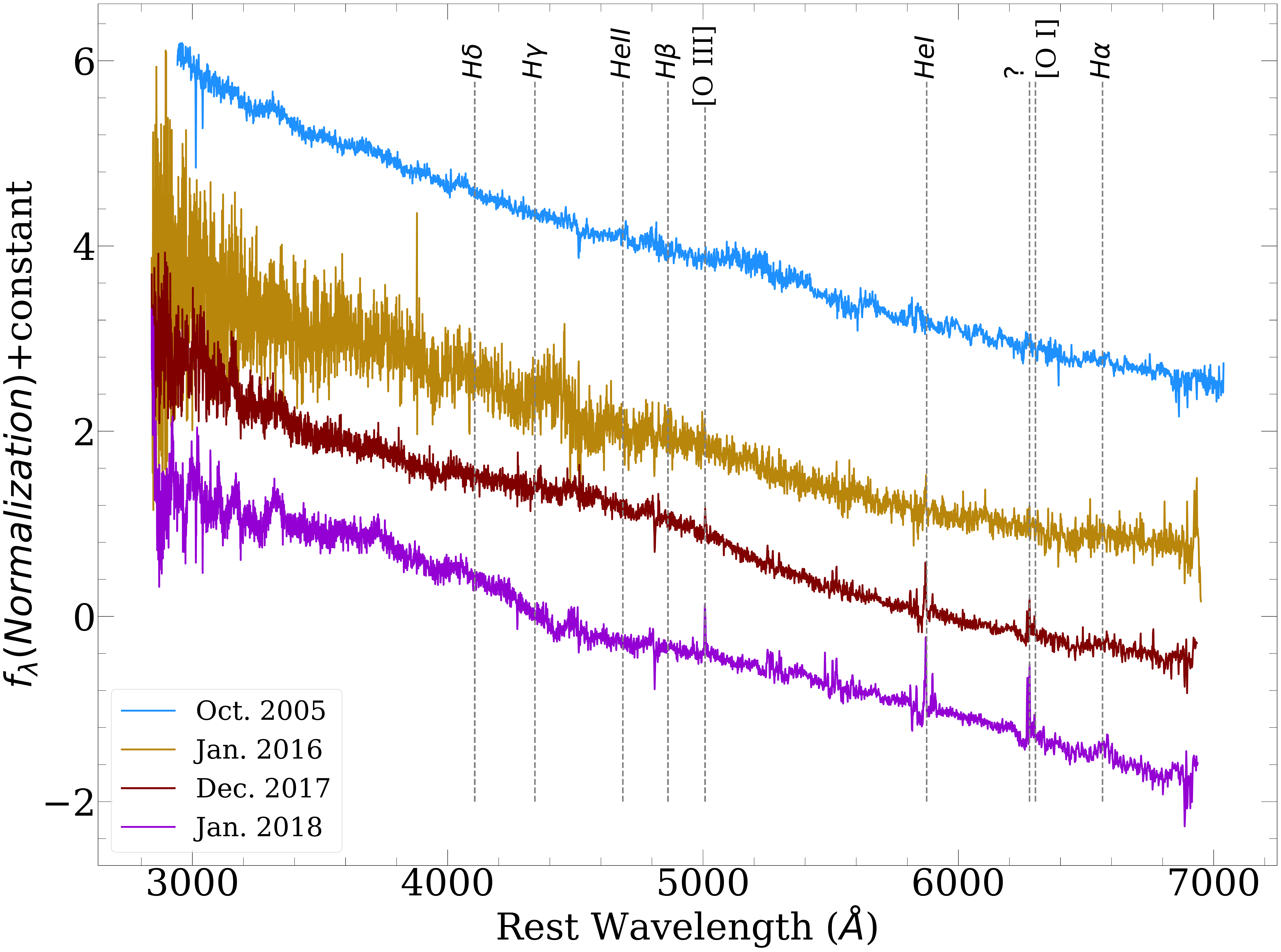}
  \caption{Optical spectra collected from four epochs. The blue line corresponds to the preoutburst spectrum extracted from SDSS, observed in 2005 October, and the dark goldenrod curve corresponds to another preoutburst spectrum which was observed in 2016 January by LAMOST.  Two postoutburst spectra extracted from LAMOST were collected in 2017 December (the maroon curve) and 2018 January (the dark violet curve). The spectra are not corrected for Milky Way extinction. The most relevant spectral lines are marked.
 }\label{fig:opt_spec}
\end{figure}

\begin{figure*}
  \centering
  \subfigure []{
  \includegraphics[width=0.4\textwidth]{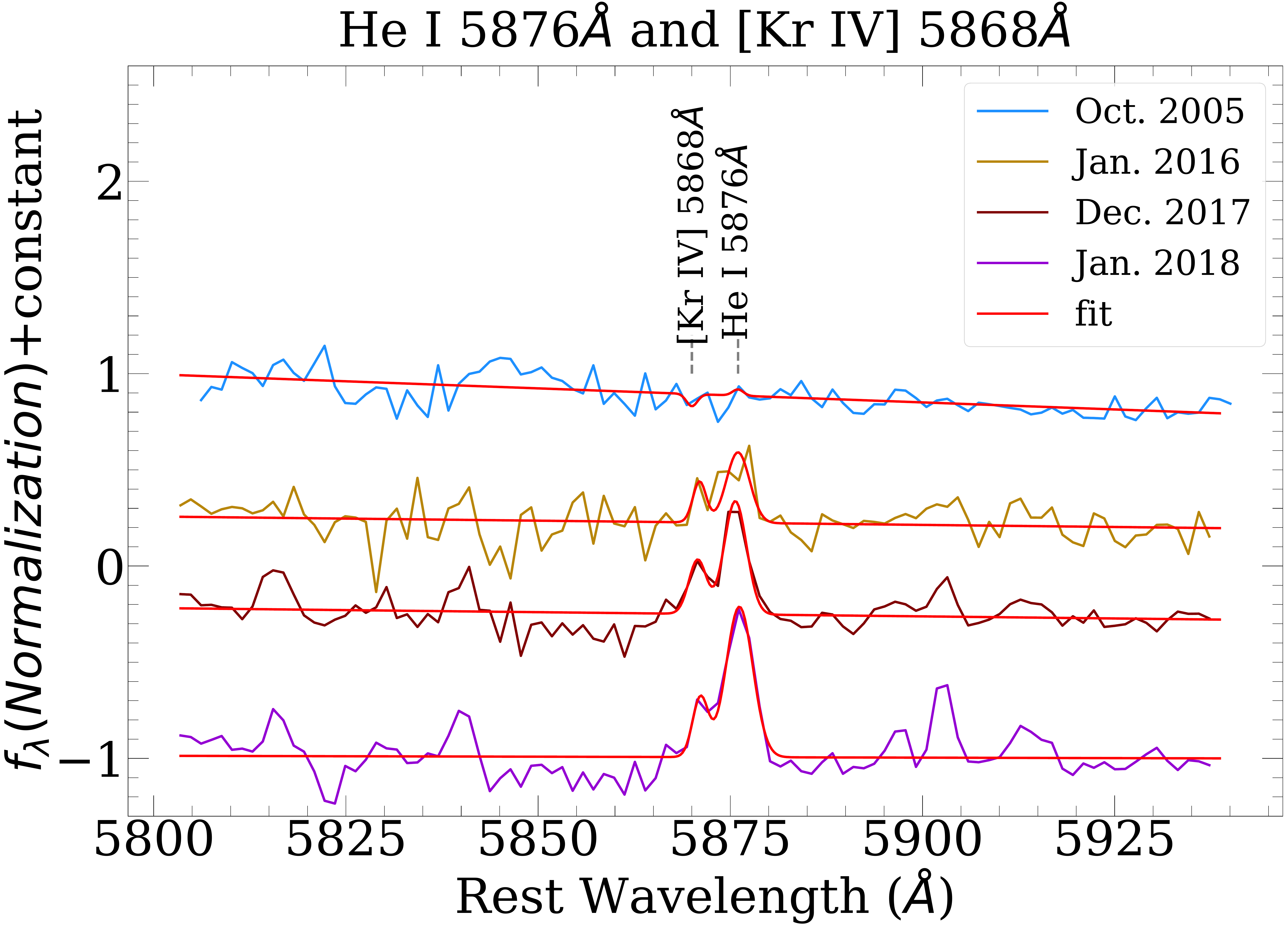}}
  \subfigure []{
  \includegraphics[width=0.4\textwidth]{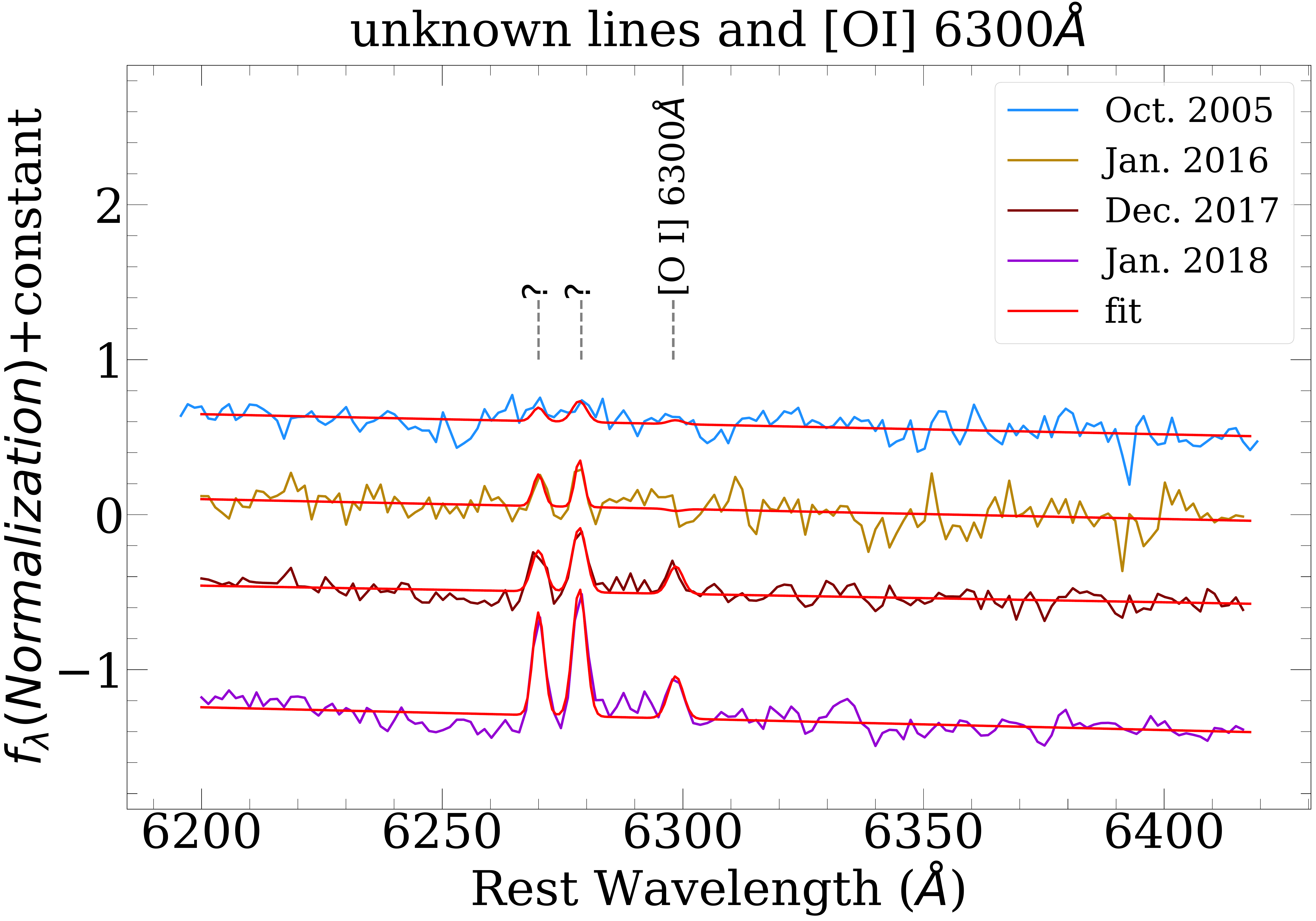}}
\subfigure []{
  \includegraphics[width=0.4\textwidth]{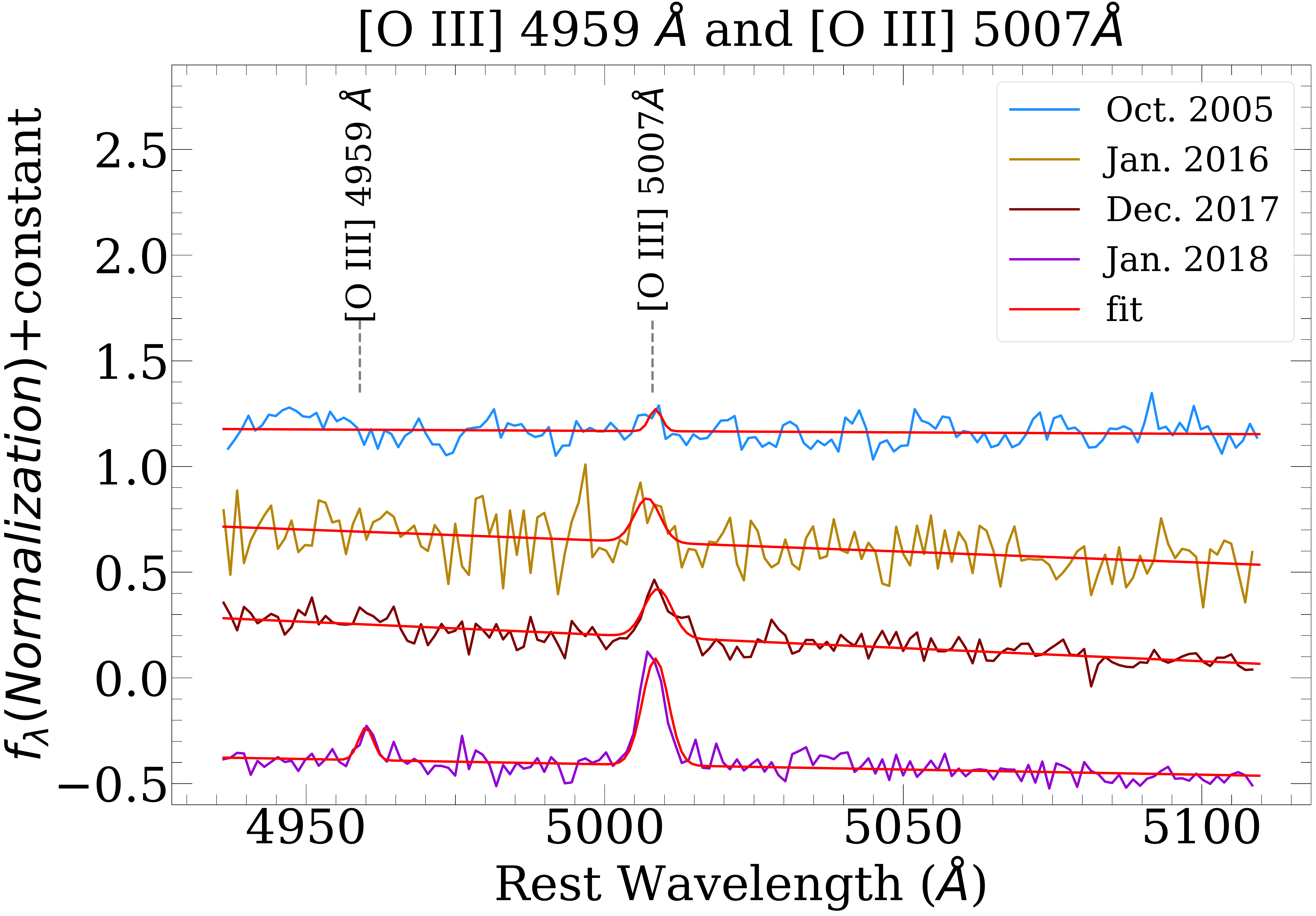}}
  \subfigure []{
  \includegraphics[width=0.4\textwidth]{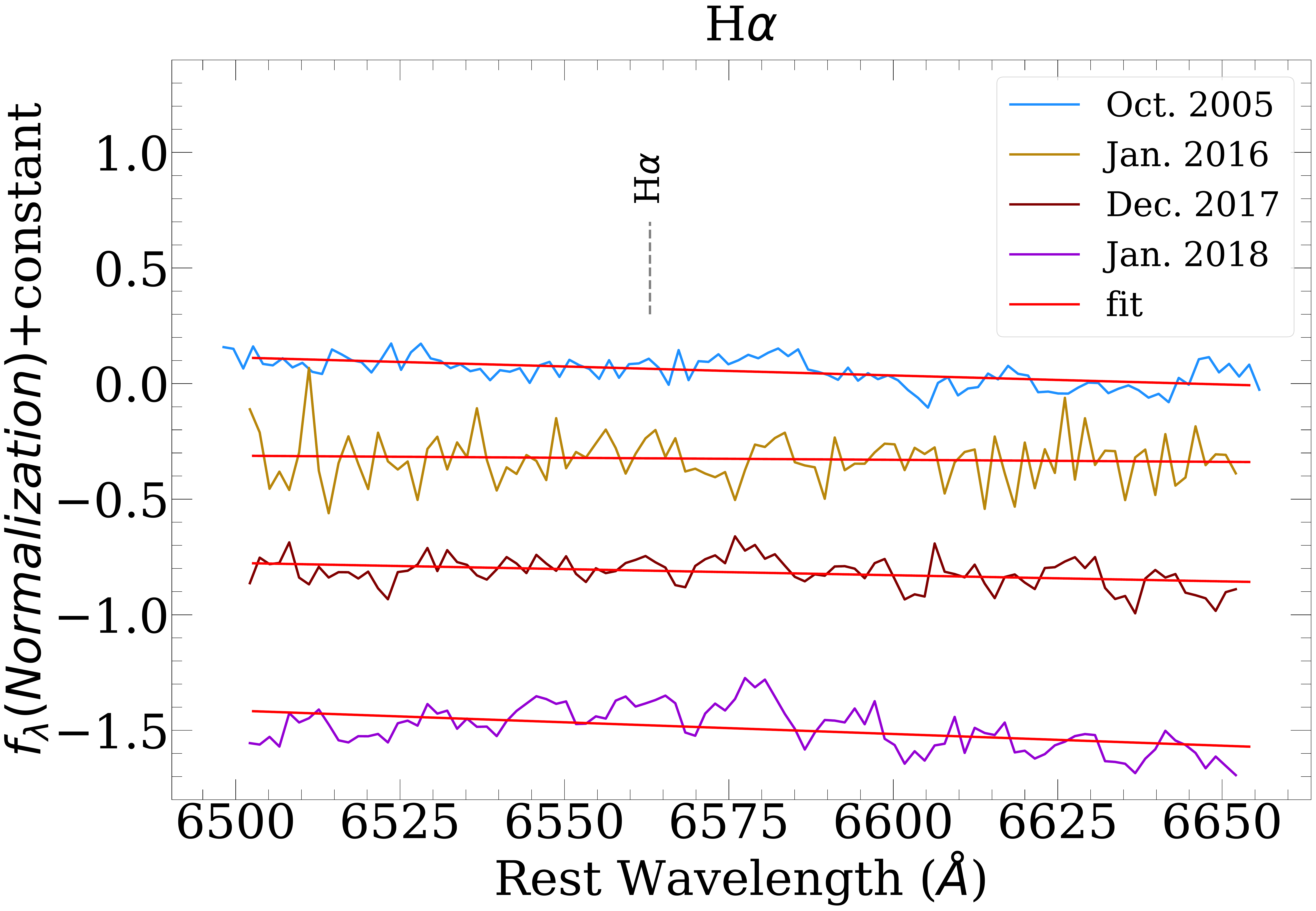}}
  \caption{Optical spectra collected from four epochs before and after the outburst. The principal emission lines are fitted with Gaussian functions.  Although the low S/Ns of the spectra preclude very precise measurements,  comparison of the oxygen, helium, and iron emission lines shows a great enhancement postoutburst. }\label{fig:spec_fit}
\end{figure*}

We analyzed optical spectra (Figure~\ref{fig:opt_spec}) obtained before the outburst (2005 October and 2016 January) and 8-9 months after the outburst (2017 December and 2018 January), which are what we could obtain from the archived spectral surveys. The spectra were shifted on wavelength according to the spectroscopic redshift $z$ = 0.306.
The postoutburst spectra exhibit a combination of narrow absorption and emission lines. The prominent emission features were detected and identified as He~I 5876 {\AA}, [O~III] 4959 {\AA}, [O~III] 5007 {\AA}, [O~I] 6300 {\AA}, and, possibly, [Kr~IV] 5868 {\AA} (see Figure~\ref{fig:spec_fit}).
Surprisingly, no hydrogen lines were detected in the spectra at any epoch. In the panel (d) in Figure~\ref{fig:spec_fit}, we demonstrate the spectra at H$\alpha$ region.
About 17 other emission lines were found in both postoutburst spectra. Two prominent emission lines at 6268 and 6276 {\AA} are presented in the panel (b) in Figure~\ref{fig:spec_fit}.
We have investigated other sources in the literature, where the optical spectra of TDEs were analyzed (e.g. \citealt{arcavi2014,leloudas2019,trakhtenbrot2019,blagorodnova2019,onori2019}). However the identification of the emission lines in our spectra is still complicated.

We define the variation of the strength of the emission lines as $\Delta I=\frac{I_t-I_i}{I_t+I_i}$, where $I_i$ is the normalized intensity of the emission line in preoutburst spectra, and $I_t$ is the corresponding intensity in the postoutburst spectra. Therefore, the error for the variation of the strength is $\sigma_I = \frac{2\sqrt{I_t^2 \sigma_i^2 + I_t^2 \sigma_t^2 }}{(I_t + I_i)^2}$, where $\sigma_t$ and $\sigma_i$ are the errors of the normalized intensity of the emission line in postoutburst and preoutburst spectra, respectively. The He I 5876 {\AA} line appears as an absorption line in 2005 while for the time after 2016, it became emission one. Apart from that, the comparison of the strength of the He I 5876 {\AA} emission line between 2016 and 2018 suggests an enhancement $\Delta I = 0.42\pm0.49$. The similar enhancement appears for unknown emission lines when we compare the spectra of 2016 and 2018, with a variation of $\Delta I = 0.56\pm0.70$ and $\Delta I = 0.60\pm0.45$, respectively. These spectral lines variation is similar to the case of PS1-10jh which shown broad helium lines in the TDE \citep{gezari2012}.
In our case, we do not detect any broad emission lines indicating the TDE, because our postoutburst spectra were obtained after 8-9 months after the end of the outburst. We attribute the detected narrow emission lines to the surrounding gas. A portion of the gas could be ejected in the surroundings after the TDE occurred.

\section{Discussion}
The outburst during 2016--2017 was reported and studied before  \citep{rakshit2017,siejkowski2017,komossa2017}, and it was explained as an aftereffect of secondary black hole impacted on disk \citep{valtonen2017,kushwaha2018,gupta2019,komossa2020,komossa2021} or a combination of shock and the jet's precession \citep{kapanadze2018}. In this section, we discuss the three origins of the outburst, including the jet's precession, aftereffect of black hole-disk impaction and the TDE. With the observation and theory, we find that a TDE may be a possible inducement for this event.

\subsection{\emph{Precession of the jet}}
It should be noted that in a binary black hole system, precession of the jet is unavoidable because of the gravitational torque of the secondary black hole on the disk of the primary one \citep{abraham2000}. A precession of the jet, which produces a variation of the angle between the line of sight of the observer and the jet and a change of Doppler factor which would lead to a quasiperiodic modulation that could be observed.

When the jet points very close to the observer, an outburst in the light curves occurs. For OJ 287, the precession period of the jet is roughly 24 yr and with the zero time of 1991.3 \citep{britzen2018}, the predicted times when the jet points directly toward Earth again should be roughly 2008.85 and 2032.85. For AGNs, the $\gamma$-rays may originate from the base of the jet through the inverse Compton (IC) process so that when the jet is pointing to the Earth, a strong $\gamma$-ray signal can be observed. However, if the $\gamma$-ray emission region is located in the broad line region, the radiation may not be observed due to the pair absorption \citep{kushwaha2018}.

\begin{figure}
  \centering
  \includegraphics[width=0.5\textwidth]{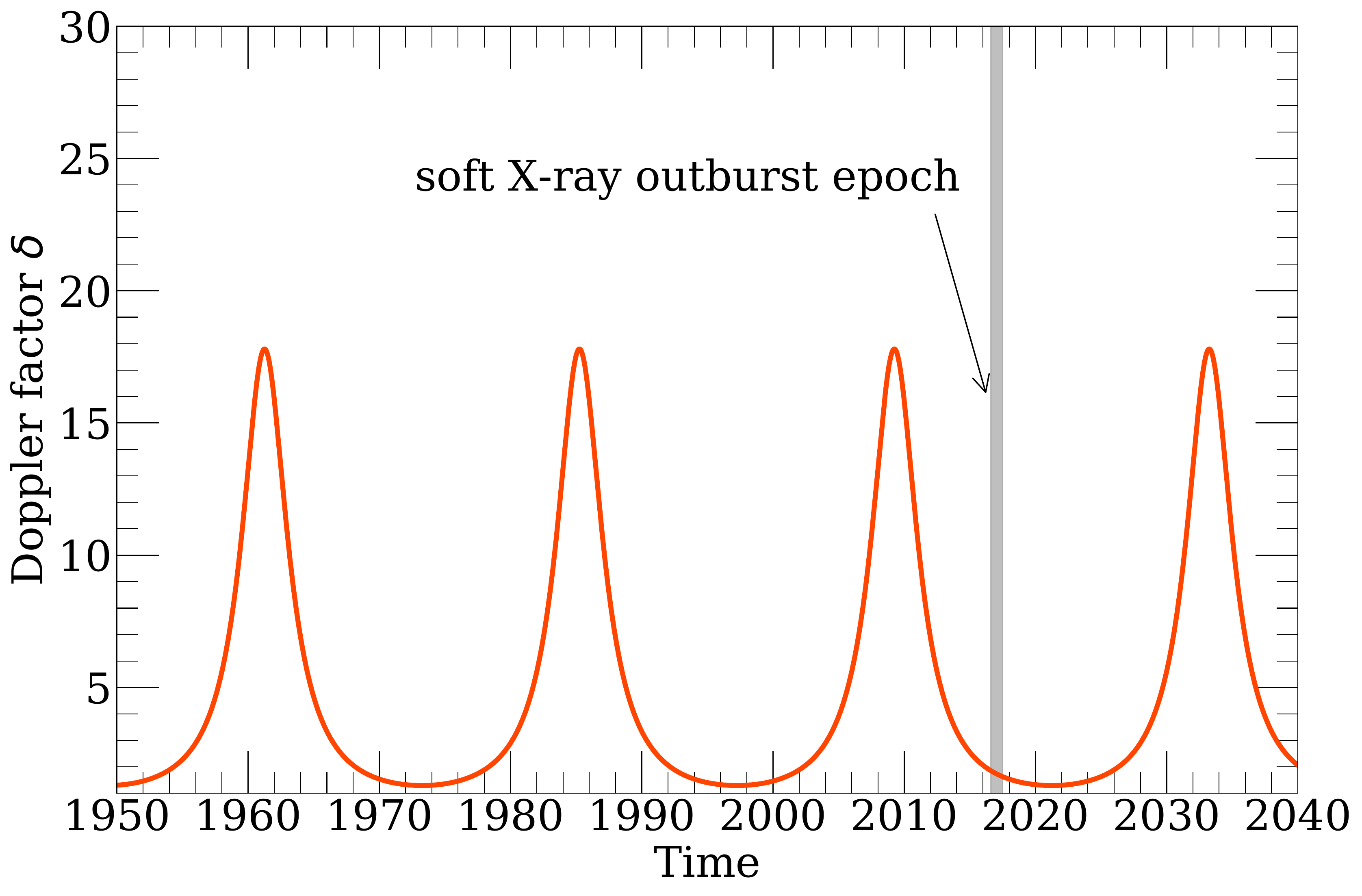}
  \caption{Simulated the evolution of the light curve in the jet precession model. The parameters are extracted from the fitted results of \cite{britzen2018}. The y-axis is the Doppler factor. The shaded area marks the X-ray outburst epoch 2016--2017. }\label{fig:pre_flux}
\end{figure}

The simulation of the light curve following the method of \cite{britzen2018} can be seen in Figure \ref{fig:pre_flux}. The parameters in the jet precession model are directly extracted from their results. For the contribution of Doppler boosting, the observed flux is related to a Doppler factor with the
form $F_{\nu}\propto \delta^{p+\alpha}$, where $p=2$ for the continuous jet and $p=3$ for the discrete case, and $\alpha$ is the spectral index \citep{britzen2018}. Here we are only concerned with the evolution of the Doppler factor which determines the profile and timing of the light curve in jet precession scenario. In our simulation result, during 2016--2017 (the shadow area in the Figure \ref{fig:pre_flux}), $\delta$ reaches the trough, which means that the flux decreases to the valley in the light curve. Furthermore, \cite{kapanadze2018} revealed a weak correlation between the $\gamma$-ray and X-ray and the $\gamma$-ray and \textsl{uvw2}. For the X-ray during the declining time, the inconspicuous time delay between the X-ray and optical/UV bands may have originated from the same spatial \citep{kushwaha2018}. In the predicted light curves of the jet precession model, the rising portion should be symmetric with the declining one, but the observed multiband light curves are not symmetric. For these reasons, we doubt that the outburst is related to the jet motion, and other origins should be considered.

\subsection{\emph{The perturbation from a black hole impacted on the disk}}
As the binary black hole system, the center of OJ 287 contains a primary black hole with the a mass $1.835\times 10^{10}M_{\sun}$ that is orbited by a secondary one with a mass $1.5\times 10^{8}M_{\sun}$ \citep{dey2018}. In the leading model for this system, the secondary black hole impacts the accretion disc around the primary twice every 12 yr. As the secondary impacts on the accretion disk, a bubble is produced, and it expands until it becomes optically thin, when the produced radiation can be seen \citep{pihajoki2016}. Hence, the optical light curves manifest the double-peaked structure seen in the 12 yr quasiperiodic outbursts \citep{lehto1996}.  When the perturbation caused by the impact event propagates in the accretion disk, the accretion rate may be affected; hence, additional fluctuations in the light curves due to the shock in the jet occur \citep{pihajoki2013,valtonen2017,kushwaha2018,kapanadze2018,komossa2020,komossa2021}.

\begin{figure}
  \centering
  \includegraphics[width=0.45\textwidth]{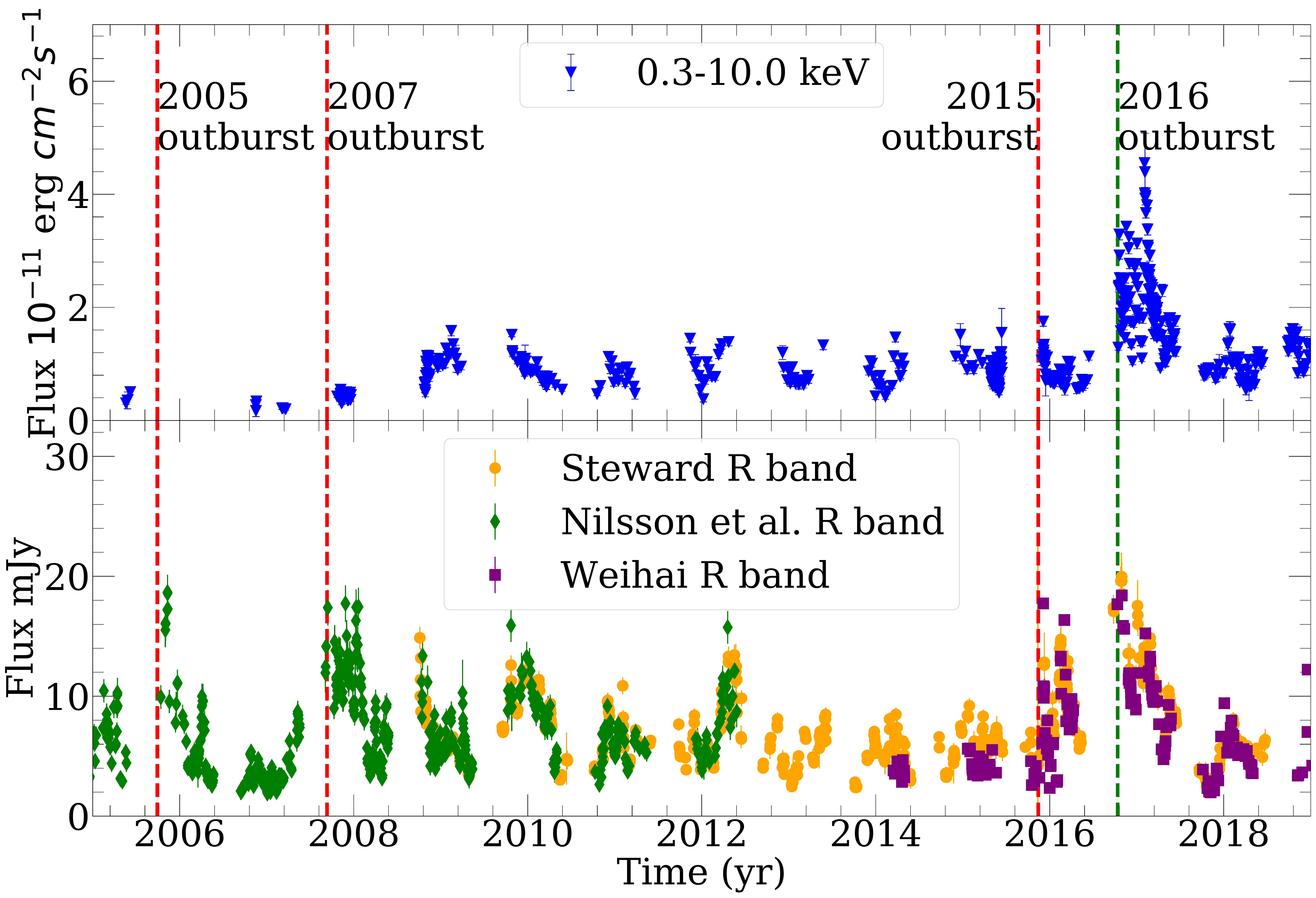}
  \caption{The X-ray and optical  light curves of OJ 287 since 2005. The dashed lines correspond to four outburst times, with the red dashed lines marking the beginnings of outbursts in 2005, 2007 and 2015, respectively, and green dashed line marking the outburst in 2016. The X-ray emission in the upper panel comes from \emph{Swift/XRT}. The historical \textsl{R} band data extracted from \cite{nilsson2018} covered 2005--2012. Most of the \textsl{R}-band data from the 2008--2019 period were observed by Steward Observatory and Weihai Observatory, respectively.}\label{fig:his_lc}
\end{figure}

The optical \textsl{R} band light curve of OJ 287 since 2005 can be seen in Figure \ref{fig:his_lc}. The peak times when outbursts are understood to occur due to the secondary impact events are marked by dashed lines; however, different from the \textsl{R} band, the distinct outburst in X-rays did not emerge until 2016 and was not clearly related to the postoutburst caused by the impact perturbation. The outbursts caused by the perturbation of black hole impact events were successfully explained by \cite{pihajoki2013}, but the X-rays are still in relatively quiescent states in those epochs. According to \cite{dey2018}, the optical outburst in 2015 was caused by the impact event in 2013 at a distance of 17566 au from the primary black hole; hence, the outburst may be related to shock in the jet produced by this event \citep{valtonen2017,kushwaha2018,kapanadze2018,komossa2020,komossa2021}. However, if the outburst started in 2016 October is related to the impact event in 2013, such similar X-ray behavior should be observed for several times in the history. Nevertheless, in the monitoring history, OJ 287 was faint in the X-ray and did not exhibit extreme outbursts before 2016 \citep{kapanadze2018}. The spectral energy distribution (SED) analyses by \cite{kushwaha2018} suggested that such a unique new component had never been seen before. However, one possibility for OJ 287 being faint in the X-ray before is that the lack of intensive X-ray observations in history may lead to missing some extreme X-ray outbursts. Moreover, the timing of the propagation of perturbation from the impact site to the jet is still unclear \citep{valtonen2017}. Therefore, the aftereffect of black hole-disk impaction could be a possibility for the X-ray outburst, and this scenario may not be completely ruled out.

\subsection{\emph{A TDE candidate in binary black holes systems?}}
The X-ray emission from AGNs located at the low redshift typically exhibits spectra well fitted by a power-law with index $\Gamma \sim 1.7$ \citep{tozzi2006,marchesi2016,auchettl2018}. \cite{auchettl2018} compared the X-ray properties between TDEs and AGNs and found that the HR of AGNs evolves with flare time and soft X-ray luminosity, but in the case of TDEs, it only shows slight variation. Contrary to the samples of AGNs and blazars, for the case of OJ 287, our results in Figure \ref{fig:hr_evo} show that a slight variation in HR with time and luminosity during the outburst. The evolution of the HR is similar to the X-ray properties of TDEs.

When a star moves too close to the black hole, it will be captured and then disrupted with roughly half of the matter bound in the form of a transient accretion disk and another portion ejected in the form of wind \citep{rees1988}. TDE, accompanied with a prominent outburst in X-ray, UV and optical bands, and the light curves usually decline as $t^{-5/3}$ including nonjetted TDEs \citep{rees1988,phinney1989,esquej2008,gezari2012,komossa2015,gezari2017,
holoien2019} and jetted TDEs \citep{bloom2011,burrows2011,levan2011,reis2012}, even in binary black holes systems \citep{liu2009,saxton2012,liu2014,shu2020}. For the outburst of OJ 287 in 2016 October -- 2017 April, the X-ray light curve rapidly rose to a prominent height level in monitored history, especially for the soft X-ray. In Figure \ref{fig:xlc}, the light curves and the HR reveal that it is an individual component in the X-ray of OJ 287. Moreover, \cite{kapanadze2018} and \cite{kushwaha2018} presented the same radiation region of the X-ray, UV, and optical bands. In addition, after the outburst, the helium and oxygen forbidden emission lines are prominently enhance. The detected narrow emission lines are associated with the surrounding gas, which could be ejected as wind after the TDE.
Combining the soft spectra, low HR and optical emission lines, we suspect that the outburst may be related to a TDE in an supermassive black hole (SMBH) system. In this subsection, we explore the possibility that the outburst was caused by the TDE in the SMBH system of OJ 287.

\subsubsection{TDE in the SMBH system}
Considering a star with mass $M_*$ and radius $R_*$, and a black hole mass $M_{BH}$, the tidal radius is
\begin{equation}
R_t = R_*\left(\frac{M_{BH}}{M_*}\right)^{\frac{1}{3}},
\end{equation}
which means that the star would be disrupted if its distance from the black hole became less than $R_t$ \citep{hills1975}. However, $R_t$ is constrained so that $R_t \geq R_{ISCO}$, where $R_{ISCO}$ is the innermost stable circular orbit (ISCO) around a black hole. In general relativity, the ISCO of a Kerr (rotating) black hole is
\begin{equation}
R_{ISCO}=\frac{GM_{BH}}{c^2}\left\{3+z_2 \mp \left[(3-z_1)(3+z_1+2 z_2)\right]^{\frac{1}{2}}\right\},
\end{equation}
where $z_1=1+\left(1-a_*^2\right)^{\frac{1}{3}}\left[(1+a_*)^{\frac{1}{3}}
+(1-a_*)^{\frac{1}{3}}\right]$, $z_2=(3a_*^2+z_1^2)^{\frac{1}{2}}$, $a_*$ is the dimensionless spin parameter of the black hole, and ``$\mp$" takes ``$-$" for clockwise spin while ``$+$" is for counterclockwise spin.
Therefore, for a Kerr black hole, the mass is limited by
\begin{equation}
M_{BH} \leq 3.16\times 10^8M_{\sun} m_*^{-\frac{1}{2}}r_*^{\frac{3}{2}}\xi^{-\frac{3}{2}},
\end{equation}
where $\xi=3+z_2 \mp \left[(3-z_1)(3+z_1+2 z_2)\right]^{\frac{1}{2}}$, $r_*=\frac{R_*}{R_{\sun}}$, and $m_*=\frac{M_*}{M_{\sun}}$. For the extreme Kerr black hole, $a_*=1$, the limit on the black hole mass is $M_{BH} \leq 3.16\times 10^8M_{\sun} m_*^{-\frac{1}{2}}r_*^{\frac{3}{2}}$ for the case of corotation and $M_{BH} \leq 1.17\times 10^7M_{\sun} m_*^{-\frac{1}{2}}r_*^{\frac{3}{2}}$ for counterrotation.
For a nonrotating Schwarzschild black hole, the mass is constrained by
$M_{BH} \leq 2.15\times10^7M _{\sun} r_*^{\frac{3}{2}}m_*^{-\frac{1}{2}}$.

The primary SMBH in OJ 287 is understood to have a very  high mass of $1.835\times 10^{10}M_{\sun}$ and an intermediate spin, $a = 0.33$ \citep{dey2018}. Hence, a main-sequence star would be swallowed rather than disrupted, and only a red giant star could possibly be disrupted.  However, for the secondary SMBH  with a mass around $1.5\times10^8M_{\sun}$, a nonspinning black hole \citep{valtonen2010} can also only disrupt a red giant or perhaps an upper main-sequence star, but if the black hole is spinning \citep{pihajoki2013}, then the much more common lower main-sequence stars are also eligible to be disrupted.

Assuming that a TDE occurred in OJ 287, with $\beta={R_t}/{R_p}$ and $M_6={M_{BH}}/{10^6 M_{\sun}}$, where $R_p$ is the pericenter of the star, the fallback rate of the debris can be written as
\begin{equation}
\dot{M}_{fb}=\frac{M_*}{T_{min}}\left(\frac{T-T_0}{T_{min}}\right)^{-5/3},
\end{equation}
where $T_0$ is the disrupt time, and the minimum fallback time $T_{min}=3.54 \times 10^6 M_6^{\frac{1}{2}}m^{-1}_*\beta^{-3}r_*^{\frac{3}{2}}~\text{s}$ \citep{rees1988,lodato2011,liu2014}. Following the method by \cite{lodato2011}, the fallback rate can be rewritten as
\begin{equation}
\dot{M}_{fb}=1.87 \times 10^{26}M_6^{-\frac{1}{2}}m_*^2\beta^{3}r_*^{-\frac{3}{2}}
\left(\frac{T-T_0}{T_{min}}\right)^{-\frac{5}{3}} \text{g}~\text{s}^{-1}.
\end{equation}

Then the luminosity can be obtained as
\begin{eqnarray}\label{equ:lumin}
L&=&\eta \dot{M}_{fb}c^2 \nonumber \\
&=& 1.71\times 10^{47}\eta M_6^{-\frac{1}{2}}m_*^{2}\beta^{3}r_*^{-\frac{3}{2}}
\left(\frac{T-T_0}{T_{min}}\right)^{-\frac{5}{3}} \text{erg}~\text{s}^{-1},\nonumber \\
\end{eqnarray}
where $\eta$ is a parameter that can be determined by observation results. Hence, if we define $t=({T-T_0})/{T_{min}}$, the evolution of luminosity in a TDE with time will be $\sim t^{-5/3}$.

\subsubsection{Light-curve fitting}

\begin{figure*}
  \centering
  \subfigure [X-ray light curves]{
  \includegraphics[width=0.45\textwidth]{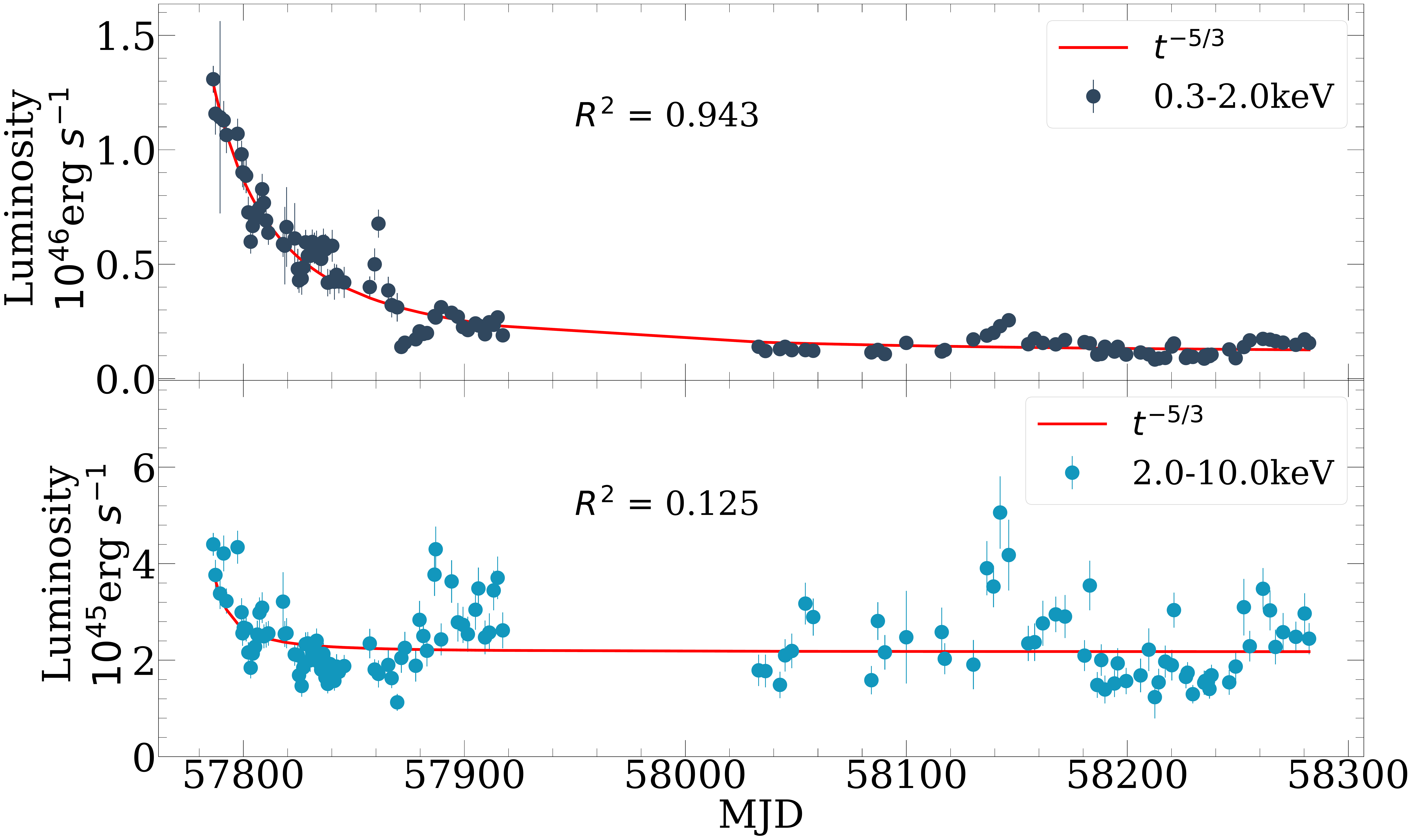}}
  \subfigure [UV-optical light curves]{
  \includegraphics[width=0.45\textwidth]{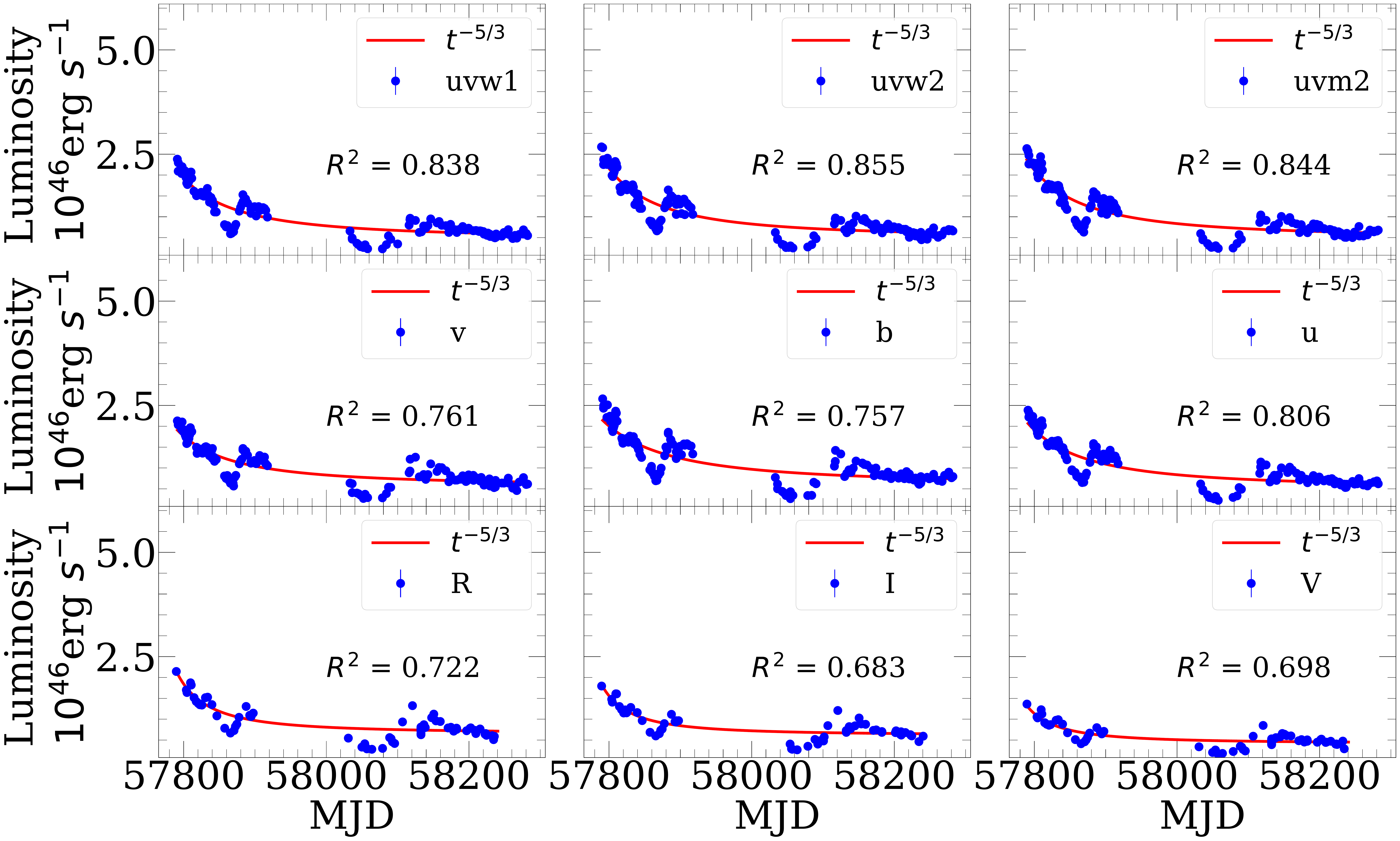}}
  \caption{The declining portions of the light curves are fitted by a power law $t^{-5/3}$. Panel (a) shows the results of the soft and hard X-rays, and the results for all nine UV-optical bands are shown in panel (b). The $R^2$ value for each fit denotes the coefficient of determination and shows excellent agreement for the soft X-ray band, as well as good agreement for the UV and optical bands.  }\label{fig:lc_fit}
\end{figure*}

\begin{figure}
  \centering
  \includegraphics[width=0.45\textwidth]{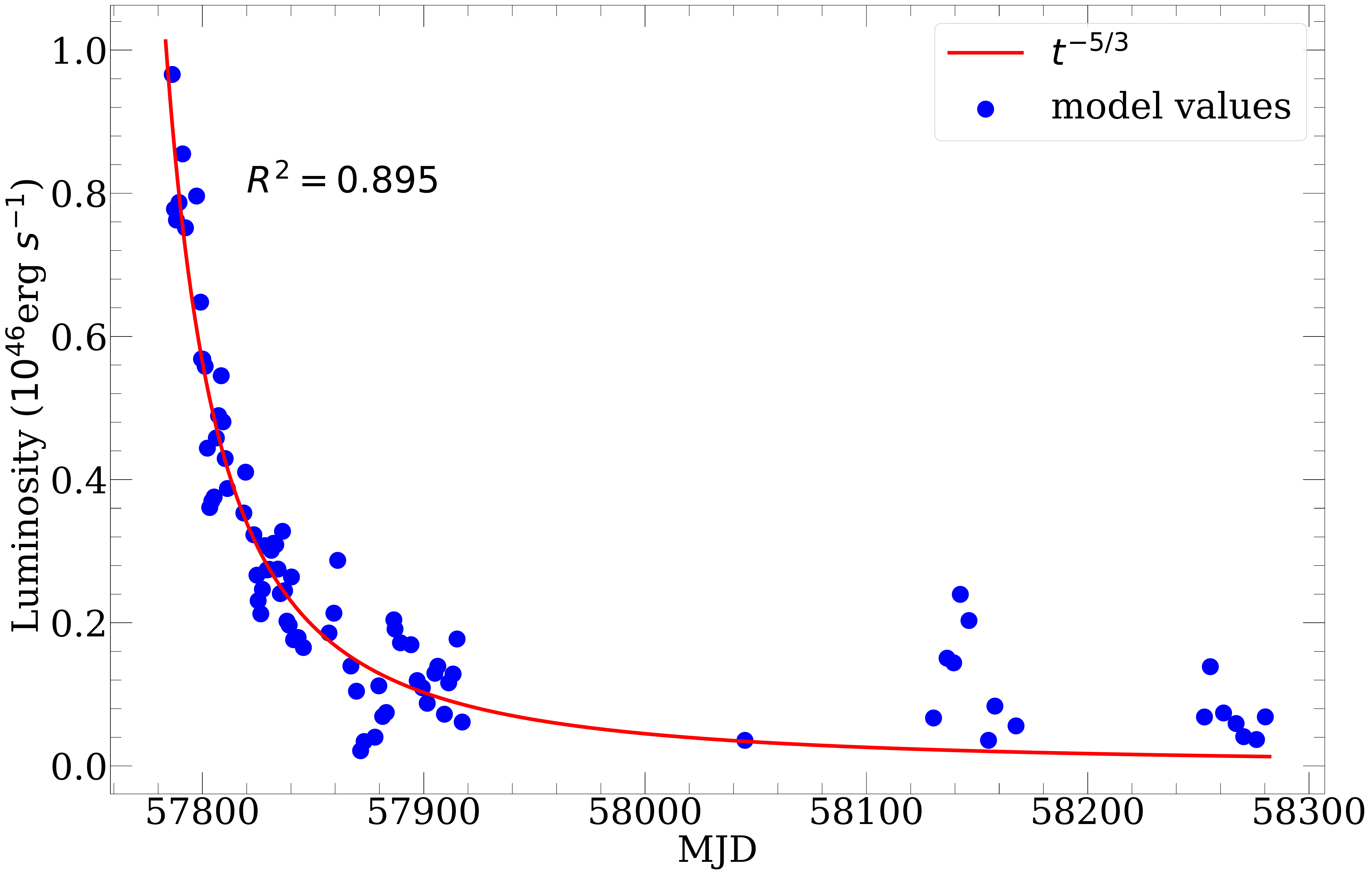}
  \caption{The values of the soft X-ray luminosity model are nicely fitted by the TDE power-law decay, $t^{-5/3}$.  }\label{fig:flux_model}
\end{figure}

\begin{deluxetable*}{cccccc}
\tablenum{2}
\tablecaption{The best-fit results for declining light curves \label{tab:lc_fit}}
\tablewidth{0pt}
\tabletypesize{\scriptsize}
\tablehead{
\colhead{Band} &  \colhead{k} & \colhead{$T_0$} & \colhead{$T_{min}$} & \colhead{h} & \colhead{$R^2$}\\
\colhead{} &  \colhead{$10^{46}~\text{erg}~\text{s}^{-1}$} & \colhead{} & \colhead{days} & \colhead{$10^{46}~\text{erg}~\text{s}^{-1}$} & \colhead{}
}
\decimalcolnumbers
\startdata
$0.3\mbox{--}2.0 \text{keV}$ & $0.36^{+0.28}_{-0.20}$ & $57741.18^{+9.59}_{-12.51}$ & $91.96^{+53.80}_{-30.17}$ & $0.11^{+0.01}_{-0.01}$ & 0.943\\
$2.0\mbox{--}10.0 \text{keV}$ & $0.01^{+0.05}_{-0.01}$ & $57773.93^{+7.90}_{-38.89}$ & $67.75^{+74.62}_{-26.84}$ & $0.22^{+0.01}_{-0.01}$ & 0.125\\
uvw1 & $4.13^{+5.49}_{-2.98}$ & $57668.86^{+38.95}_{-55.11}$ & $70.18^{+72.04}_{-34.16}$ & $0.47^{+0.07}_{-0.08}$ & 0.838\\
uvw2 & $4.21^{+5.43}_{-3.10}$ & $57677.25^{+34.10}_{-53.53}$ & $70.32^{+72.09}_{-33.28}$ & $0.49^{+0.07}_{-0.09}$ & 0.855\\
uvm2 & $4.44^{+5.22}_{-3.24}$ & $57669.18^{+38.05}_{-54.95}$ & $71.75^{+71.11}_{-33.17}$ & $0.49^{+0.07}_{-0.09}$ & 0.844\\
v & $4.05^{+5.55}_{-3.14}$ & $57664.64^{+47.94}_{-57.53}$ & $65.17^{+76.86}_{-33.76}$ & $0.57^{+0.07}_{-0.08}$ & 0.761\\
b & $7.62^{+11.64}_{-5.93}$ & $57636.15^{+52.34}_{-34.10}$ & $59.53^{+80.61}_{-30.08}$ & $0.60^{+0.09}_{-0.08}$ & 0.757\\
u & $6.77^{+12.36}_{-5.47}$ & $57656.10^{+45.22}_{-50.43}$ & $55.90^{+85.73}_{-30.31}$ & $0.51^{+0.08}_{-0.08}$ & 0.806\\
R& $5.69^{+13.20}_{-5.19}$ & $57719.70^{+36.11}_{-86.21}$ & $31.25^{+95.81}_{-19.32}$ & $0.66^{+0.06}_{-0.09}$ & 0.722\\
I & $4.68^{+13.88}_{-4.34}$ & $57725.29^{+34.86}_{-89.62}$ & $28.61^{+97.72}_{-17.77}$ & $0.61^{+0.05}_{-0.07}$ & 0.683\\
V & $4.59^{+13.74}_{-4.31}$ & $57719.67^{+37.10}_{-87.47}$ & $26.31^{+102.87}_{-15.63}$ & $0.42^{+0.04}_{-0.06}$ & 0.698\\
\enddata
\tablecomments{From the X-ray to optical bands, we fitted the decline light curves ranging from MJD 57786 to 58300 with $t^{-5/3}$. }
\end{deluxetable*}

According to the theory of TDEs, the light curve should decline with time $t$ as a relation $t^{-5/3}$ which is thereby taken to be a special characteristic of such events \citep{rees1988,phinney1989}. The decay portions of the light curves of OJ 287 during the outburst epoch were fitted by the relation $L_t=k \left(\frac{T-T_0}{T_{min}}\right)^{-5/3}+h$ using the Markov Chain Monte Carlo (MCMC) code \emph{emcee} \citep{foreman2013}, where $L_t$ is the total luminosity and $h$ corresponds to the background luminosity in OJ 287 and the results of the X-ray fits are shown in Figure \ref{fig:lc_fit}. The fitted parameters of the soft X-ray light curve are $k=0.36^{+0.28}_{-0.20}\times 10^{46}~\text{erg}~\text{s}^{-1}$, $T_0=\text{MJD}~57741.18_{-12.51}^{+9.59}$, the $T_{min}=91.96_{-30.17}^{+53.80}~\text{days}$ and the background luminosity $h=0.11^{+0.01}_{-0.01}\times 10^{46}~\text{erg}~\text{s}^{-1}$.  The coefficients of determination ($R^2$) in the fitting are 0.943 and 0.125 for the soft X-ray and the hard X-ray, respectively. So the light curve of the soft X-ray was very well fitted by this TDE shape, while the hard X-ray was not, which may not surprising because the hard X-rays are more likely related to the jet. The light curves from \emph{Swift} monitoring show that during the outburst, the flux of the hard X-rays was influenced only slightly and this behavior is consistent with the $\gamma$-ray and radio light curves. More details of the best-fit results for the X-ray and UV/optical are shown in Table \ref{tab:lc_fit}.

 We model the luminosity with the results from the spectrum fitting. The fitting results of the second redshift power law in the model DzPL were input into \texttt{xspec 12.10.1} when we used a single zPL model to describe the new component; then, luminosity can be obtained. With this approach, the model luminosity is fit by $L_{model}=k t^{-5/3}$, and the result is shown in Figure \ref{fig:flux_model}; we obtain $T_0=\text{MJD} ~57743.8$ and $T_{min}=89.5~\text{days}$, with the $R^2=0.895$. At late times, it appears that modest additional flares are present but are presumably independent of the TDE.

 It should be noted that the decline of the luminosity from modeling can be well explained by a standard power law $t^{-5/3}$. The fitting results are also close to those from the light curve of the soft X-ray directly.

\subsubsection{Explanation of the TDE}

The results of the fits in Figure \ref{fig:lc_fit} reveal that the light curves of the UV and optical bands also roughly follow the $t^{-5/3}$ expectation, but not as strictly as for the soft X-rays, presumably because there are greater contributions from the preexisting accretion disk or the jet in other bands. Before the X-ray outburst, the flux of the UV and optical bands stayed at a high level, presumably due to the secondary black hole impact to the primary SMBH disk \citep{valtonen2016}; therefore, it is difficult to distinguish the contribution of a TDE. For a TDE in an AGN, the collision between the debris stream and a preexisting accretion disk may produce fast outflows that can accelerate relativistic electrons through a bow shock \citep{chan2019,xu2019}. As a result, at the beginning (around MJD 57681) of the outburst era, the synchrotron emission may contribute to the radiation; thus, the high polarization degree accompanying the high luminosity at the early phase of the outburst can be seen (see the bottom panel of Figure \ref{fig:mwlc}).

During the outburst, the times for the maximum flux for the optical/UV and X-ray are MJD 57681 and MJD 57786, respectively. The observations reveal that the maximum flux time for the X-ray lags the optical/UV for 105 days. Similar lags are seen in the TDEs ASASSN-15oi \citep{gezari2017} and AT 2019azh \citep{liu2020}. The collision of the fallback debris from the disrupted star would cause the outburst of the optical/UV bands \citep{piran2015}, whereas in the early epoch of a TDE, the outflow from disruption may obscure the inner disk, and as a result, the X-rays are absorbed and converted to optical/UV emission \citep{metzger2016}. Another possibility is that inefficient circularization delays the accretion, which would also contribute to the X-ray peak lagging behind the optical/UV peak \citep{gezari2017}. With the estimation of the circularization timescale $t_{circ}=340.3~ M_6^{-\frac{7}{6}}\beta^{-3}~\text{day}$, where $\beta$ is the penetration factor \citep{gezari2017}, we prefer the first possibility. For the case of a blazar hosted by an SMBHB system, the process may be much more complex.

The spectra of the new component of the X-ray can be nicely described by the second redshift power law. A similar case was found in XMMSL2 J144605.0+685735, whose spectra were well fitted by a single power law because of the thermal photons upscattered by the warm electrons \citep{saxton2019}. During the outburst, the flux of the radio bands showed the smooth enhancement. However, radio observation can also be detected in a TDE, for example, ASASSN-14li \citep{alexander2016,velzen2016}. The radio emission in the TDE may originate from the unbound debris, relativistic jets and subrelativistic outflows \citep{roth2020}. These possibilities may explain the radio observation in this outburst of OJ 287.

\subsubsection{The progenitor star in the TDE}
The helium-rich TDE PS1-10jh was first reported when very weak hydrogen lines and strong helium emission lines were seen in its spectra \citep{gezari2012}. Subsequent work agreed that a red giant was disrupted in TDE PS1-10jh \citep{bogdanovic2014}, arguing that the hydrogen envelope was stripped when a red giant passed by the black hole and the helium core was disrupted only after the hydrogen envelope was accreted by the black hole. However, the case has also been made that a main-sequence star was disrupted in PS1-10jh \citep{guillochon2014}. In their opinion, the hydrogen envelope may be ionized and so produce few emission lines. The postoutburst optical spectra of OJ 287, which exhibit strong helium and oxygen emission lines and very weak hydrogen lines, may be related to either of the above two mechanisms. Therefore, we cannot exclude either one of the progenitors and both a main-sequence star and a red giant should be considered.

Assuming that a solar-type star with $m_*=r_*=1$ is disrupted by the secondary black hole,  by substituting the fitting results into Equation (\ref{equ:lumin}) and considering the red shift of $z=0.306$, then we have $\eta_{\text{0.3-2.0keV}}= 0.040$ and $\beta=1.761$. Alternatively, assuming a red giant with mass $m_*=1.5$ and radius $r_*=100$ \citep{macleod2012,bogdanovic2014},  we find $\beta=15.386$ and $\eta_{\text{0.3-2.0keV}}=0.027$ for the case of the disruption in the vicinity of the secondary black hole.  For such a red giant, disruption around the rotating primary SMBH is also possible, yielding the parameters  $\beta=34.275$ and $\eta_{\text{0.3-2.0keV}}=0.027$.

We recall that the Eddington luminosity of an SMBH is
\begin{equation}
L_{Edd} = 1.25\times 10^{46} \left(\frac{M}{10^8M_\odot}\right) \text{erg} ~\text{s}^{-1}.
\end{equation}
So, for the primary black hole of OJ 287, the Eddington luminosity would be $\sim 10^{48} \text{erg} ~ \text{s}^{-1}$, while for the case of the secondary black hole,  $L_{Edd}\sim 10^{46}\text{erg} ~ \text{s}^{-1}$. Since the super-Eddington accretion would occur in the early epoch of the TDE and make a comparison with the observed peak luminosity of $\sim 10^{46} \text{erg}~\text{s}^{-1}$ \citep{kushwaha2018}, we find that the star was more likely disrupted by the secondary black hole.

Although the peak luminosity of $\sim 10^{46}\text{erg}/\text{s}$ is rare in discovered TDEs, it reaches the Eddington luminosity of the secondary black hole of OJ 287 (black hole mass is $1.5\times 10^8 M_\sun$). In the sample in \cite{auchettl2017}, most of the TDE candidates reveal a peak luminosity of $< 10^{46}\text{erg}/\text{s}$, because most of them have a black hole mass of $< 10^8 M_\sun$. However, in their sample, three TDE candidates exhibit the peak luminosity $> 10^{46}\text{erg}/\text{s}$. These three candidates reveal the power law decay in light curve, hard X-ray spectra, high HR, and luminosity much higher than Eddington luminosity indicating the signature of the jet. Besides, \cite{auchettl2017} presented a positive relation between the peak X-ray luminosity and redshift in their TDE sample. And we find that this soft X-ray luminosity in the outburst event and the redshift of OJ 287 are consistent with this result.

Finally, we list the observed phenomena involving the outburst of 2016--2017 in Table \ref{tab:tde_jet}. Some of these features may be explained by the TDE or jet activity, and we list whether they can be explained by these two scenarios.

\begin{deluxetable}{ccc}
\tablenum{3}
\tablecaption{Summary of the observed properties and whether They Supports the TDE or jet activity \label{tab:tde_jet}}
\tablewidth{0pt}
\tabletypesize{\scriptsize}
\tablehead{
\colhead{Observed Property} & \colhead{TDE} &  \colhead{Jet Activity}
}
\decimalcolnumbers
\startdata
Softer-when-brighter & Yes & Yes \\
Evolution of HR& Yes & no \\
Evolution of emission lines & Yes & no \\
$t^{-5/3}$ declining light curves & Yes & no \\
High luminosity & Yes & Yes \\
High optical polarization & Yes & Yes\\
\enddata
\tablecomments{Summary of the properties for this outburst event. We list whether the phenomenon can be explained by the scenario.}
\end{deluxetable}

\section{Conclusion}
We investigate the origin of the outburst of OJ 287 in 2016--2017 through the light curves, X-ray energy spectra, HR and optical spectra. Three possibilities, including the jet precession, aftereffect of black hole-disk impaction and the TDE, are discussed in the previous sections.

During 2016 October--2017 April, an extremely prominent outburst in the soft X-ray was observed. The outburst also occurred in the wavelength from the X-ray, UV, and optical bands, but it did not show in the $\gamma$-ray. In addition, at the beginning and end of the outburst, a high polarization degree can be observed, but in the X-ray flux peak time, the polarization degree reached a lower level. The above results are consistent with the previous works \citep{komossa2017,kapanadze2018,kushwaha2018,komossa2020,komossa2021}. During the outburst, the HR was located at an extremely low level in the monitoring history, and it evolved insignificantly with time and soft X-ray luminosity. After the outburst, the prominent narrow helium and oxygen emission lines were detected in the optical spectra. We attribute the appearance of these lines to the surrounding gas, which could be ejected due to the TDE. The multiwavelength light curves of OJ 287 and the results of the correlation analysis from \cite{kapanadze2018} and \cite{kushwaha2018} suggest the same radiation region of the X-ray, UV, and optical bands during the interval from MJD 57786 to 57869, but not the $\gamma$-ray and radio bands. The ``softer-when-brighter" phenomenon reveals a new component during the X-ray outburst. This behavior disagrees with the prediction of the jet precession model. Shock in the jet produced by the black hole-disk impaction may induce the outburst at that time. However, the comparison of the historical light curves between the \textsl{R} band and the X-ray shows that the timing of the optical outburst before this event did not match the case of the X-ray; therefore, the soft X-ray outburst may not be related to this impaction. Nevertheless, the simultaneity of the light curves in the X-ray, UV, and optical bands, and the ``softer-when-brighter" behavior of X-ray, negligible evolution of the HR with flare time and soft X-ray luminosity, and the enhancement of helium and oxygen emission lines suggest that the new component may be produced by the TDE. The light curves, including the optical, UV, and X-ray bands, are well fitted by the TDE declining power law $t^{-5/3}$, which reaches agreement with the prediction by TDE. The disrupted star is probably disrupted by the secondary black hole, and its progenitor may be a red giant or a main-sequence star; to reach an exact interpretation, further research is needed. In addition, the light curve of the soft X-ray agrees with the theoretical prediction of the TDE in the SMBHB system. Finally, we argue that the outburst detected in the SMBHB system OJ 287  in 2016 October--2017 April period may be explained with the TDE.

Recently, a second prominent X-ray outburst from 2020 April to 2020 June was observed. The ``softer-when-brighter" behavior was shown in the X-ray and the outburst covering the X-ray to optical was reported \citep{komossa2020}. The SED of this outburst exhibited a new component that is similar to the case of 2016--2017, but with a different correlation between the X-ray and optical/UV bands in the later time \citep{kushwaha2020}. The study of the outburst in 2020 is beyond the scope of this work; we are preparing another work for this event.

If our interpretation is correct, this event would be the first TDE that has been observed in an SMBHB system of AGNs. It supplies some important implications for the study of TDEs in complicated systems and on the range of mechanisms producing light-curve variations for AGNs. In addition, statistical studies of such TDEs in blazars could provide key information for understanding of the growth of black holes and their coevolution with galaxies. Moreover, a TDE found in such an SMBHB implies the role of the general relativity effect, and the Newtonian gravity may not describe the event well.

\acknowledgments

We thank Professor Paul J. Wiita for some fruitful suggestions for our work.  We also thank Dr. Stefanie Komossa for the discussion of this work. We also thank Professor Fukun Liu, Dr. Rossi Elena Maria, Professor Richard Saxton, and Professor Rongfeng Shen for the valuable advice on TDE theory, and Professor Mauri Valtonen and Dr. Pihajoki Pauli provide precious advice for the theory of OJ 287. This work is supported by the Natural Science Foundation of China under grant No.~11873035, the Natural Science Foundation of Shandong province (No.~JQ201702), and the Young Scholars Program of Shandong University (No.~20820162003). S. Alexeeva acknowledges support from the National Natural Science Foundation of China (grant No.~12050410265) and LAMOST FELLOWSHIP program that is budgeted and administrated by Chinese Academy of Sciences.

We acknowledge Swift for the observation and providing the public data for the research.

Data from the Steward Observatory spectropolarimetric monitoring project were used. This program is supported by Fermi Guest Investigator grants NNX08AW56G, NNX09AU10G, NNX12AO93G, and NNX15AU81G. This research has made use of data from the OVRO 40-m monitoring program \citep{richards2011} which is supported in part by NASA grants NNX08AW31G, NNX11A043G, and NNX14AQ89G and NSF grants AST-0808050 and AST-1109911.

Guoshoujing Telescope (the Large Sky Area Multi-Object Fiber Spectroscopic Telescope, LAMOST) is a National Major Scientific Project built by the Chinese Academy of Sciences. Funding for the project has been provided by the National Development and Reform Commission. LAMOST is operated and managed by the National Astronomical Observatories, Chinese Academy of Sciences.

Funding for the Sloan Digital Sky Survey IV has been provided by the Alfred P. Sloan Foundation, the U.S. Department of Energy Office of Science, and the Participating Institutions. SDSS-IV acknowledges
support and resources from the Center for High-Performance Computing at
the University of Utah. The SDSS web site is \url{www.sdss.org}.

SDSS-IV is managed by the Astrophysical Research Consortium for the
Participating Institutions of the SDSS Collaboration including the
Brazilian Participation Group, the Carnegie Institution for Science,
Carnegie Mellon University, the Chilean Participation Group, the French Participation Group, Harvard-Smithsonian Center for Astrophysics,
Instituto de Astrof\'isica de Canarias, The Johns Hopkins University, Kavli Institute for the Physics and Mathematics of the Universe (IPMU) /
University of Tokyo, the Korean Participation Group, Lawrence Berkeley National Laboratory,
Leibniz Institut f\"ur Astrophysik Potsdam (AIP),
Max-Planck-Institut f\"ur Astronomie (MPIA Heidelberg),
Max-Planck-Institut f\"ur Astrophysik (MPA Garching),
Max-Planck-Institut f\"ur Extraterrestrische Physik (MPE),
National Astronomical Observatories of China, New Mexico State University,
New York University, University of Notre Dame,
Observat\'ario Nacional / MCTI, The Ohio State University,
Pennsylvania State University, Shanghai Astronomical Observatory,
United Kingdom Participation Group,
Universidad Nacional Aut\'onoma de M\'exico, University of Arizona,
University of Colorado Boulder, University of Oxford, University of Portsmouth,
University of Utah, University of Virginia, University of Washington, University of Wisconsin,
Vanderbilt University, and Yale University.

\software{astropy \citep{astropy2013}, emcee \citep{foreman2013}, matplotlib \citep{hunter2007}. }

\end{document}